\documentclass[showpacs,prd,nofootinbib,reprint,preprintnumbers]{revtex4-1}
\usepackage{adjustbox}
\usepackage[utf8]{inputenc}
\usepackage{color,graphicx,epsfig}
\usepackage{amsmath,amssymb,dsfont,epsfig,graphicx,xcolor,fontenc}
\usepackage{bm}
\usepackage[english]{babel}
\usepackage{graphicx}%
\usepackage{amsfonts}%
\usepackage{amssymb}
\usepackage{braket}
\usepackage{hyperref}
\usepackage{epstopdf}
\usepackage{ulem}
\usepackage{soul}
\usepackage{float}
\usepackage{multirow}

\newcommand{\mc}{\mathcal}

\def\LjubljanaFMF{Faculty of Mathematics and Physics, University of Ljubljana,
 Jadranska 19, 1000 Ljubljana, Slovenia }
\def\LjubljanaIJS{Jo\v zef Stefan Institute, Jamova 39, 1000 Ljubljana, Slovenia}
\def\IFIC{Instituto de F\'isica Corpuscular,
Universitat de Val\'encia – Consejo Superior de Investigaciones Cient\'ificas,
Parc Cient\'ific, E-46980 Paterna, Valencia, Spain}

\begin{document}

\title{Impact of new invisible particles on $B\to K^{(*)} E_{\rm miss}$ observables}

\author{Patrick D. Bolton}
\email[Electronic address: ]{patrick.bolton@ijs.si} 
\affiliation{\LjubljanaIJS}

\author{Svjetlana Fajfer}
\email[Electronic address: ]{svjetlana.fajfer@ijs.si} 
\affiliation{\LjubljanaIJS}
\affiliation{\LjubljanaFMF}

\author{Jernej~F.~Kamenik}
\email[Electronic address: ]{jernej.kamenik@cern.ch} 
\affiliation{\LjubljanaIJS}
\affiliation{\LjubljanaFMF}

\author{Mart\'in Novoa-Brunet}
\email[Electronic address: ]{martin.novoa@ific.uv.es}
\affiliation{\IFIC}

\begin{abstract}
Motivated by a recent Belle~II measurement that suggests an excess in the rare decay $B \to K\, E_{\rm miss}$, and building upon our recent differential decay rate likelihood analysis of the existing experimental information, we investigate possible new physics (NP) scenarios in which light invisible states participate in flavour-changing  $b \to s$ transitions. In particular, we consider the total and  differential $B\to K^* E_{\rm miss}$ decay rates and $K^*$ polarisation effects in each NP scenario preferred by the $B\to K E_{\rm miss}$ measurement. We show that future measurements of these $B \to K^* E_{\rm miss}$ observables will offer decisive discrimination among the different NP explanations. Our results highlight the strong complementarity of the rare semi-invisible $b$-hadron decay observables, and underline the importance of analysing their momentum transfer spectra when
probing extensions of the Standard Model that feature new light degrees of freedom. 
\end{abstract}

\maketitle

%%%%%%%%%%%%%%%%%%%%%%%%%%%%%%%%%%%%%%%%%%
%
\section{Introduction}
\label{sec:intro}
%
%%%%%%%%%%%%%%%%%%%%%%%%%%%%%%%%%%%%%%%%%%

The $b$-hadron decays to (semi-)invisible final states, $B\to K^{(\ast)} E_{\rm miss}$, $B_s\to E_{\rm miss}$, $B_s\to \phi E_{\rm miss}$ and $\Lambda_b \to\Lambda   E_{\rm miss}$, offer challenging tests of the Standard Model (SM). 
The semi-invisible processes are mediated in the SM by the well-known flavour-changing neutral current (FCNC) transition $b\to s\nu\bar\nu$~\cite{Buras:2014fpa,Buras:2020xsm,Becirevic:2023aov}, with the missing energy $E_{\text{miss}}$ originating from an undetected pair of massless neutrinos. Tantalisingly, the most precisely measured branching ratio at present is
\begin{align}
\label{eq:BelleII-result}
{\cal B}(B^+\to K^+ E_{\rm miss}) = (2.3\pm 0.7) \times 10^{-5}\,,
\end{align}
from the Belle~II collaboration~\cite{Belle-II:2023esi}, lying $2.9\sigma$ above the SM prediction~\cite{Parrott:2022zte}. In the $b\to s \nu \bar \nu$ interpretation, the $B\to K E_{\rm miss}$ decay is only sensitive to the vector part of the weak quark current ($\bar b \gamma_\mu s$), while the decay $B \to K^{\ast} E_{\text{\rm miss}}$ receives contributions from both vector and axial-vector ($\bar b \gamma_\mu \gamma_5 s$) currents. This dependence can provide complementary information on the underlying short-distance dynamics, given that the relevant hadronic matrix elements are well understood within existing theoretical frameworks~\cite{Becirevic:2023aov, Gubernari:2023puw}.
Currently, only experimental upper bounds apply for the branching ratio $B\to K^{\ast} \,E_{\rm miss}$, the most stringent being 
\begin{align}
\label{eq:Belle-result}
\mathcal{B}(B^0 \to K^{*0}E_{\text{\rm miss}}) < 1.8 \times 10^{-5}~(90\%~\text{CL})\,,
\end{align}
from the Belle collaboration~\cite{Belle:2017oht}. Nonetheless, it already puts important constraints on possible explanations of the $B\to K E_{\rm miss}$ discrepancy~\cite{Bause:2023mfe,Allwicher:2023xba}. In addition, an upper bound on the invisible branching fraction of $B_s$ decays,
\begin{align}
\label{eq:ALEPH-result}
{\cal B}(B_s \to E_{\rm miss}) < 5.6 \times 10^{-4} ~(90\%~\text{CL}) \,,
\end{align}
has been reported in Ref.~\cite{Alonso-Alvarez:2023mgc}, based on a recast of ALEPH data~\cite{ALEPH:2000vvi}. Recently, we found that this mode can also put competitive bounds on particular interpretations of the Belle II result~\cite{Bolton:2024egx}.

In the wake of the intriguing experimental situation, numerous new physics (NP) explanations of the discrepancy between the Belle II measurement and the SM prediction have been suggested, with some based on explicit NP model constructions (see, e.g., Refs.~\cite{Berezhnoy:2025tiw, Lin:2025jzp,Lee:2025jky,Hu:2024mgf, Bhattacharya:2024clv, Buras:2024mnq, Altmannshofer:2024kxb, Becirevic:2024iyi,Allwicher:2024ncl,Hati:2024ppg, Kim:2024tsm, Becirevic:2024pni, DAlise:2024qmp, He:2024iju,Chen:2024cll,Gabrielli:2024wys, Loparco:2024olo,Ho:2024cwk,McKeen:2023uzo, Altmannshofer:2023hkn,Datta:2023iln,Berezhnoy:2023rxx, Chen:2023wpb,He:2023bnk,Wang:2023trd, Felkl:2023ayn,Athron:2023hmz}) and others parametrising possible NP effects in the effective field theory (EFT) framework (see, e.g., Refs.~\cite{Guedes:2024vuf,Buras:2024ewl,Rosauro-Alcaraz:2024mvx,Marzocca:2024hua,Hou:2024vyw,Fridell:2023ssf}). In general, there are two distinct possibilities to account for the missing decay rate. One is based on the assumption that only SM neutrinos contribute to $E_{\rm miss}$ in the final state (see e.g. Refs.~\cite{Allwicher:2023xba,Bause:2023mfe}), while the other one allows for new undetected particles in addition to the SM neutrinos (see e.g. Ref.~\cite{Bolton:2024egx}). In the first case, existing experimental constraints require the presence of a right-handed quark current ($\bar b \gamma_\mu (1+\gamma_5) s$) contribution~\cite{Allwicher:2023xba,Bause:2023mfe}. In the second case this requirement can be avoided, essentially because the amplitudes involving the SM neutrinos and NP invisible states do not interfere. 

Building upon previous work~\cite{Kamenik:2009kc}, we have recently explored the landscape of general NP scenarios with additional light invisible fields coupling to $(\bar b \Gamma s)$ currents (where $\Gamma \in \{ 1, \gamma_5, \gamma_\mu, \gamma_\mu \gamma_5, \sigma_{\mu\nu} \}$), making use of the EFT approach~\cite{Bolton:2024egx}. We have systematically considered both single scalar and vector particle contributions to the $B\to K E_{\rm miss}$ final state, as well as production of pairs of scalars, spin 1/2 and 3/2 fermions, and vectors. In addition to the relevant $B\to K^{(*)} E_{\rm miss}$ and $B_s\to E_{\rm miss}$ branching ratios, we have included all available kinematical distributions presented by the Belle II and BaBar collaborations to construct likelihoods for various possible invisible final states. Consequently, we have discerned viable NP scenarios and determined the corresponding invisible particle masses and EFT parameters preferred by the current data.  

With its final dataset, Belle~II is expected to  measure $\mathcal{B}(B\to K^{(*)} E_{\rm miss})$ to unprecedented precision below $10^{-7}$~\cite{Belle-II:2018jsg}. This could allow for detailed (differential) kinematical studies of both decay modes. Since the presence of new light invisible particles in the final state can significantly alter the expected distributions, such studies are particularly relevant in NP scenarios with new light degrees of freedom.  

Based on our previous results and in anticipation of upcoming Belle II measurements, in the present work we explore in more detail possible NP effects in $B\to K^\ast E_{\rm miss}$ decays. We focus on the differential decay rates and $K^*$ polarisation effects in each of the NP scenarios preferred by the $B\to K E_{\rm miss}$ measurement. In particular, we highlight important correlations between both decay modes, as well as their complementarity due to different kinematics, spin and hadronic (i.e. form factor) effects. 

The remainder of this article is structured as follows: in Section~\ref{sec:model}, we outline the light new fields and their effective interactions considered in this work. In Section~\ref{sec:fit} we give the results of the global fit performed in Ref.~\cite{Bolton:2024egx} to the Belle~II, BaBar and ALEPH data. In Section~\ref{sec:results}, we then explore the predictions for $B\to K^*E_{\text{miss}}$ observables, namely the differential and total decay rate and longitudinal polarisation fraction, for each favoured NP scenario. We explore the implications for future measurements at Belle~II and conclude in Section~\ref{sec:conclusions}.

%%%%%%%%%%%%%%%%%%%%%%%%%%%%%%%%%%%%%%%%%%
%
\section{Model Considerations}
\label{sec:model}
%
%%%%%%%%%%%%%%%%%%%%%%%%%%%%%%%%%%%%%%%%%%

In Ref.~\cite{Bolton:2024egx}, we studied the impact of additional invisible final states on the process $B^+ \to K^+ E_{\text{miss}}$. In particular, we found that the presence of a light scalar ($\phi$), fermion ($\psi$) or vector ($V_\mu$) field can provide a better fit to the Belle~II data than the SM prediction for $B^+ \to K^+\nu\bar{\nu}$, rescaled by the signal strength $\mu = 5.4\pm 1.5$~\cite{Belle-II:2023esi}.

For simplicity, the light new fields are assumed to be SM gauge singlets, but may transform under a dark gauge or global symmetry. If charged under some dark $U(1)$, the scalar and vector fields are complex and the fermion field is of Dirac type. If neutral, the scalar and vector fields can also be real ($\phi = \phi^*$ and $V_\mu = V_\mu^*$) and the fermion field can be of Majorana type ($\psi = \psi^c$), with the corresponding particles and antiparticles being identical. 

\begin{figure}[t!]
  \centering
  \includegraphics[width=0.49\columnwidth]{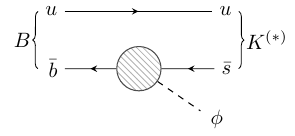}
  \includegraphics[width=0.49\columnwidth]{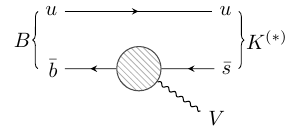}
  \includegraphics[width=0.49\columnwidth]{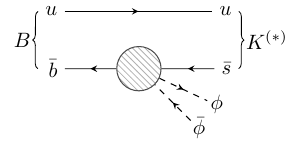}
  \includegraphics[width=0.49\columnwidth]{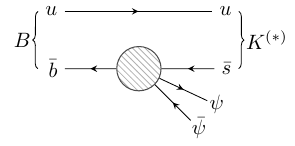}
  \caption{Invisible light new states contributing to the process $B^+ \to K^{+}E_{\text{miss}}$ which can provide a better fit to the Belle~II (ITA) data compared to the $\mu \times \text{SM}$. The same states contribute to the associated process $B \to K^* E_{\text{miss}}$, with upcoming measurements expected from Belle~II.}
  \label{fig:NP-scenarios}
\end{figure}

With these scenarios in mind, we considered in Ref.~\cite{Bolton:2024egx} the impact of $B^+ \to K^+ \sum X$ with the invisible final states
\begin{align}
\label{eq:final_states}
\sum X \in \big\{\phi, V, \phi\phi, \phi\bar{\phi}, \psi\psi, \psi\bar{\psi}\big\}\,,
\end{align}
on the $B^+ \to K^+ E_{\text{miss}}$ signal at Belle~II, as depicted in Fig.~\ref{fig:NP-scenarios}. The two-body decays require the scalar and vector fields to be singlets under any unbroken dark gauge group, while the three-body decays are compatible with the scalar and fermion fields transforming in arbitrary representations. We do not consider here the scenario with two invisible vector bosons in the final state, as it cannot improve the fit to the Belle~II data with respect to the rescaled SM ($\mu \times \text{SM}$).

As they are singlets under the SM, the light new fields can couple to any gauge-invariant product of SM fields. The renormalisable \textit{portal} interactions at dimension-four and below ($d \leq 4$) are the most promising avenues to produce them in laboratory experiments or as viable dark matter candidates. However, to contribute to the process $B^+ \to K^+ E_{\text{miss}}$, the light invisible states must couple to the flavour-changing $b \to s$ quark current. We take the EFT approach and assume that other, heavy degrees of freedom can be integrated out of the theory, resulting in non-renormalisable interactions of the light invisible states and the quarks below the electroweak scale.

In the previous study~\cite{Bolton:2024egx}, we considered the lowest dimension ($d\leq 6$) operators in the \textit{parity} basis, coupling the light invisible states to $P$-even $(\bar b \Gamma s)$ and -odd $(\bar b \Gamma \gamma_5 s)$ $b \to s$ transition quark currents with the Dirac structure $\Gamma \in \{1, \gamma_\mu, \sigma_{\mu\nu}\}$. This basis choice has the advantage of simplifying the analytical expressions for the branching fractions, reviewed in App.~B of Ref.~\cite{Bolton:2024egx}, as the $B\to K^{(*)}$ form factors are most conveniently defined via hadronic matrix elements of definite parity. However, it is more natural from the top-down perspective to write operators in the \textit{chiral} basis, which couples the light invisible states to $(\bar b \Gamma P_X s)$, where $P_X$ denotes the usual chirality projection operator with $X \in \{L,R\}$. If the heavy NP is well above the electroweak scale, operators in the weak effective theory arise from operators invariant under the SM gauge group, distinguishing left- and right-handed quarks. We adopt the chiral basis in the following.

For completeness, we present here the operators which can improve the fit to the Belle~II data~\cite{Belle-II:2023esi} compared to the $\mu \times \text{SM}$. Firstly, the effective Hamiltonian involving the vector $b \to s$ transition quark current is
\begin{align}
\mathcal{H}_{\text{eff}}^{V} &\supset (\bar{b}\gamma_\mu P_X s)\bigg[C_{dV}^{V,X}V^\mu + \frac{C_{d\phi}^{V,X}}{\Lambda}\partial^\mu \phi + \frac{C_{d\phi\phi}^{V,X}}{\Lambda^2} i \phi^* \overset{\leftrightarrow~\,}{\partial^\mu} \phi \nonumber\\
& \hspace{6em}  + \frac{C_{d\psi}^{V,XY}}{\Lambda^2}(\bar{\psi}\gamma^\mu P_Y \psi) \bigg]  + \text{h.c.}\,,
\label{eq:vector_quark}
\end{align}
for $X, Y \in \{L,R\}$. The $B^+\to K^+ V$ decay rate induced by $C_{dV}^{V,X}\neq0$ naively diverges in the $m_V\to 0$ limit, which can be cured if $V_\mu$ is associated with an active gauge symmetry. However, the available Belle~II data favours $m_V \neq 0$, so this requirement can be relaxed. For a real scalar field, the coefficient $C_{d\phi\phi}^{V,X}$ vanishes identically. For a Majorana fermion field, the relation
\begin{align}
C_{d\psi}^{V,XL} = -C_{d\psi}^{V,XR} \,,
\end{align}
holds as a consequence of the vanishing vector current.

The relevant effective Hamiltonian involving the scalar $b \to s$ transition quark current is
\begin{align}
\mathcal{H}_{\text{eff}}^{S} &\supset (\bar{b} P_X s)\frac{v}{\sqrt{2}}\bigg[\frac{C_{d\phi}^{S,X}}{\Lambda}\phi + \frac{C_{d\phi\phi}^{S,X}}{\Lambda^2}|\phi|^2 \bigg] + \text{h.c.} \,.
\label{eq:scalar_quark}
\end{align}
After integration by parts and insertion of the quark field equations of motion, it can be shown that
\begin{align}
i \frac{v}{\sqrt{2}} C_{d\phi}^{S,X} = m_b C_{d\phi}^{V,X} - m_s C_{d\phi}^{V,Y} \,,
\label{eq:phi_coupling}
\end{align}
for $X \neq Y$. While we explore the impact of $C_{d\phi}^{S,X}$ in the following, the results can easily be translated to the derivative couplings using Eq.~\eqref{eq:phi_coupling}. The normalisation of the $C_{d\phi\phi}^{S,X}$ coefficient is taken to be the same for a real or complex scalar field. A prefactor of 1/2 is typically included to account for the two possible contractions of the real scalar field in the amplitude. 

Finally, we consider the effective dipole-like operator involving the tensor $b \to s$ transition quark current,
\begin{align}
\mathcal{H}_{\text{eff}}^{T} &\supset \frac{v}{\sqrt{2}}\frac{C_{dV}^{T,X}}{\Lambda^2}(\bar{b} \sigma_
{\mu\nu}P_X s)V^{\mu\nu}  + \text{h.c.} \,,
\label{eq:tensor_quark}
\end{align}
with the field strength tensor $V_{\mu\nu} = \partial_\mu V_\nu - \partial_\nu V_\mu$, which is manifestly invariant under any dark gauge group. The scalar and tensor quark currents above are multiplied by the factor $v/(\sqrt{2}\Lambda)$, as they can only be generated by operators in the SM-invariant EFT with the $SU(2)_L$ indices of the quark and Higgs doublets contracted.

%%%%%%%%%%%%%%%%%%%%%%%%%%%%%%%%%%%%%%%%%%
%
\section{Fit Procedure}
\label{sec:fit}
%
%%%%%%%%%%%%%%%%%%%%%%%%%%%%%%%%%%%%%%%%%%

The Belle~II collaboration performed a search~\cite{Belle-II:2023esi} for $B^+\to K^+\nu\bar{\nu}$ in a $362 ~\text{fb}^{-1}$ sample of $e^+e^-$ collisions at the $\Upsilon(4S)$ resonance. Two methods were used to search for the rare decay. The inclusive tagging analysis (ITA), used inclusive properties of the $B$ meson not decaying via the signal process, $B^+\to K^+ E_{\text{miss}}$, whereas the hadronic tagging analysis (HTA) selected only for hadronic decays. The former method improves the signal efficiency at the price of a higher background, separated into $B^+B^-$, $B^0\bar{B}^0$ and continuum components. In the full and signal-rich signal regions, defined by the parameter $\eta(\text{BDT}_2)$, the ITA observed a significant excess of events above the expected background in the range of reconstructed momentum transfer $3< q^2_{\text{rec}}/\text{GeV}^2 < 5$. The two $q^2_{\text{rec}}$ bins in this range are the main drivers of the 2.9$\sigma$ tension. The HTA saw a smaller excess over the SM, with a signal strength of $\mu = 2.2_{-2.0}^{+2.4}$ and a significance of 0.6$\sigma$. Taken together, these results are an intriguing hint of an additional contribution to $B^+\to K^+ E_{\text{miss}}$. Until now, the Belle, Belle II and BaBar experiments have only been able to set upper limits on the process~\cite{BaBar:2010oqg,BaBar:2013npw,Belle:2013tnz,Belle:2017oht,Belle-II:2021rof}.

The additional invisible final states, $\sum X$, coupled via Eqs.~\eqref{eq:vector_quark}--\eqref{eq:tensor_quark}, cannot interfere with the SM and therefore can only increase the total rate for $B^+\to K^+ E_{\text{miss}}$. The predicted momentum transfer $q^2$ distribution of the enhanced rate can vary significantly between the different NP scenarios. In the SM, the process $B^+ \to K^{+}\nu\bar{\nu}$ is well understood and induced predominantly at short-distances, leading to the differential decay rate,
\begin{align}
\label{eq:dGdq2_BtoK_SM}
\frac{d\Gamma(B \to K \nu\bar{\nu})_{\text{SM}}}{dq^2} = \frac{|\vec{p}_{K}|^3}{64 \pi^3} \big|C_{d\nu}^{V,LL}\big|^2 f_{+}^2 \,,
\end{align}
with the effective coefficient,
\begin{align}
\label{eq:CdvLL}
C_{d\nu}^{V,LL} = \frac{4G_F\lambda_t }{\sqrt{2}}\frac{X_t}{s_w^2}\frac{\alpha(M_Z)}{2\pi} \,,
\end{align}
where $G_F = 1.1663788(6) \times 10^{-5}$~$\text{GeV}^{-2}$ is the Fermi constant~\cite{ParticleDataGroup:2024cfk}, $s_w^2 = 0.23129(4)$ is the sine squared of the weak mixing angle~\cite{ParticleDataGroup:2024cfk}, $\alpha(M_Z) = 1/127.930(8)$ is the $\overline{\text{MS}}$ scheme QED fine structure constant evaluated at the $Z$ boson mass~\cite{ParticleDataGroup:2024cfk}, $\lambda_t \equiv V_{tb}V_{ts}^*$ is the appropriate product of Cabbibo-Kobayashi-Maskawa (CKM) matrix elements, and $X_t =  1.469(17)$ accounts for next-to-leading-order QCD~\cite{Buras:2005gr} and two-loop electroweak~\cite{Brod:2010hi} corrections. For the relevant CKM matrix elements, we use the value found by combining the HPQCD lattice QCD calculation of the $B_s$ meson mixing matrix element~\cite{Dowdall:2019bea} with the measured $B_s$ oscillation rate~\cite{ParticleDataGroup:2024cfk}, $|\lambda_t| = 41.85(93) \times 10^{-3}$. We note that this value is larger than the $|\lambda_t|$ derived indirectly via CKM unitarity with $|V_{cb}|$ from tree-level inclusive and exclusive $B$ decays~\cite{Buras:2014fpa,Becirevic:2023aov}. The vector $B\to K$ form factor, $f_+$, is computed using the BSZ parametrisation~\cite{Bharucha:2015bzk} fit results of Ref.~\cite{Gubernari:2023puw}, with HPQCD~\cite{Bouchard:2013eph,Parrott:2022rgu} and FNAL+MILC~\cite{Bailey:2015dka} lattice QCD results as input at high $q^2$. Integrating the differential rate in Eq.~\eqref{eq:dGdq2_BtoK_SM} over the entire kinematically allowed $q^2$ range gives the SM branching fraction, 
\begin{align}
\label{eq:B_BtoK_SM}
\mathcal{B}(B^+ \to K^+ \nu\bar{\nu})_{\text{SM}} = (4.90 \pm 0.17 \pm \delta)\times 10^{-6} \,,
\end{align}
where we have used $\tau_{B^+} = 1.638(4)$~ps and divided the uncertainties into hadronic and parametric components, respectively, with the latter given by $\delta = 0.25$. For $B^+\to K^+\nu_\tau\bar{\nu}_\tau$, there is also the long-distance tree-level contribution~\cite{Kamenik:2009kc} with an intermediate $\tau$ lepton, with a size $\sim 10\%$ of Eq.~\eqref{eq:B_BtoK_SM}. This contribution is nevertheless subtracted from the final Belle~II result.

\begin{figure}[t!]
  \centering
  \includegraphics[width=0.9\columnwidth]{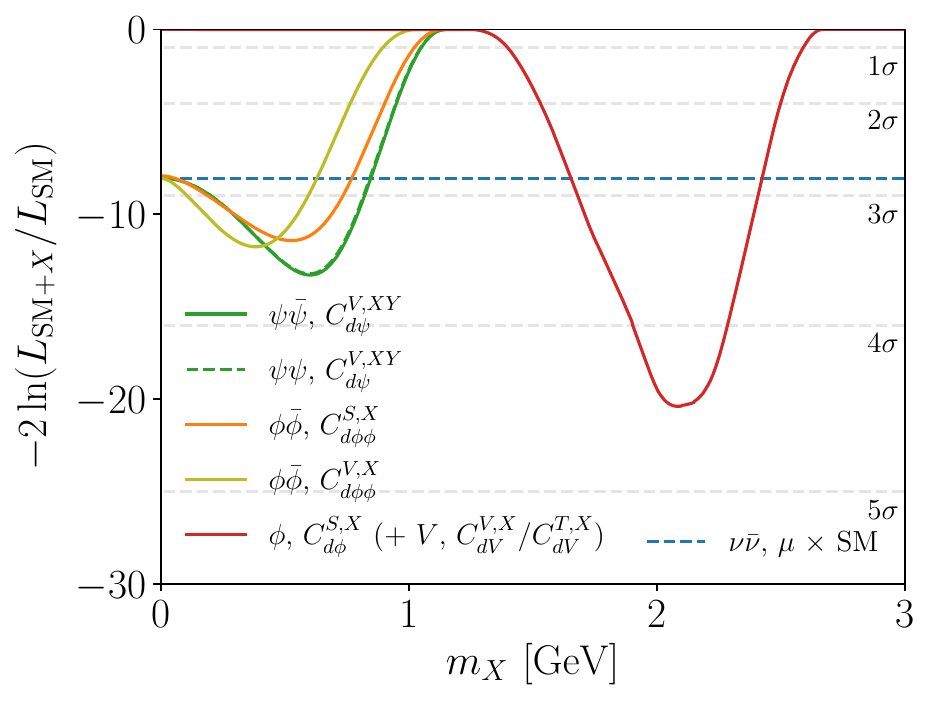}
  \caption{Minimised log-likelihood ratio of the $\mu \times \text{SM}$ and SM plus $\sum X$ scenarios with respect to the SM only hypothesis, for different values of the mass $m_{X}$.}
  \label{fig:mass-fit}
\end{figure}

The (differential) $B^+\to K^+ \sum X$ decay rates induced by the NP interactions in Eqs.~\eqref{eq:vector_quark}--\eqref{eq:tensor_quark} are provided in full in App.~\ref{app:decays}. For conciseness, we express the results in terms of the parity basis NP coefficients, which can be readily transformed to the chiral basis. The NP contributions depend on the scalar and tensor $B\to K$ form factors, $f_0$ and $f_T$, respectively. For the two-body decays, the momentum transfer is fixed to the mass of the light scalar or vector boson, $q^2 = m_{X}^2$. The three-body decays produce continuous $q^2$ distributions distinct from the SM prediction above the production threshold $q^2 = 4m_X^2$. In the limit $m_\psi \to 0$, the $q^2$ distribution for the invisible fermion pair interacting via the chiral vector operator in Eq.~\eqref{eq:vector_quark} tends towards the SM shape. The $q^2$ distribution for the invisible scalar pair interacting via the chiral scalar operator in Eq.~\eqref{eq:scalar_quark} is flatter compared to the SM, with a more pronounced fall off close to the kinematic threshold $q^2 = (m_B - m_K)^2$.

As the NP contributions modify the shape of the $q^2$ distribution, it is not possible to perform a simple recast of the best-fit branching fraction in Eq.~\eqref{eq:BelleII-result} onto the mass ($m_X$) and coupling ($C_X$) of the invisible field. The fit to the signal strength performed in Ref.~\cite{Belle-II:2023esi} assumes the $q^2$ distribution of SM $B^+ \to K^+ \nu\bar{\nu}$, and is therefore biased towards NP models that lead to a universal shift in the rate, such as the invisible fermion scenario in $m_\psi \to 0$ limit. Consequently, a full shape analysis of the Belle~II data must be performed for general model inference.

To perform the shape analysis, we require the number of expected signal events in each $q^2_{\text{rec}}$ bin in the SM and SM + $\sum X$ scenarios. First, we calculate the differential event rate in $q^2_{\text{rec}}$ as
\begin{align}
\label{eq:exp_sig}
\frac{dN}{dq^2_{\mathrm{rec}}} = N_{B\bar{B}} \int dq^2 f_{q^2_\mathrm{rec}}(q^2) \epsilon(q^2)\frac{d\mathcal{B}}{dq^2}\,,
\end{align}
where $N_{B\bar{B}} = (387 \pm 6) \times 10^{6}$ is the number of produced $B\bar{B}$ pairs, $f_{q^2_\mathrm{rec}}(q^2)$ accounts for the smearing of $q_{\text{rec}}^2$ with respect to the true momentum transfer $q^2$, and $\epsilon(q^2)$ is the detector efficiency. The binned smearing and efficiency maps were made available for the ITA by the Belle~II collaboration~\cite{Belle-II:2023esi, Belle-II:PrivateCommunication}, which we smoothly interpolate to approximate the functions in Eq.~\eqref{eq:exp_sig}. The expected signal in each bin, $s_i$, is then found by integrating Eq.~\eqref{eq:exp_sig} over the range $[q^2_{\mathrm{rec},i},q^2_{\mathrm{rec},i+1}]$.

\begin{table}[t!]
\centering
\renewcommand{\arraystretch}{1.4}
\setlength\tabcolsep{1.3pt}
\begin{tabular}{c|c|c|cc}
\hline

\multirow{2}{*}{$C_X$} & \multirow{2}{*}{$\sum X$} & \multirow{2}{*}{$m_X$/GeV} & \multicolumn{2}{c}{Best-fit ($\Lambda = 10$~TeV)}\\\cline{4-5}

 &  &  & Belle~II only & + BaBar + LEP \\ \hline

\multirow{2}{*}{$C_{d\psi}^{V,XY}$} & $\psi\bar{\psi}$ & \multirow{2}{*}{$0.60_{-0.14}^{+0.11}$} & \multirow{2}{*}{$4.9 \pm 0.7$} & \multirow{2}{*}{$4.8 \pm 0.7$} \\

 & $\psi\psi$ & & & \\ \hline
 
\multirow{2}{*}{$C_{d\phi\phi}^{S,X}$} & $\phi\bar{\phi}$ & \multirow{2}{*}{$0.53^{+0.14}_{-0.18}$} & \multicolumn{2}{c}{$(1.1 \pm 0.2) \times 10^{-1}$} \\

 & $\phi\phi$ &  & \multicolumn{2}{c}{$(7.5 \pm 1.1) \times 10^{-2}$} \\
 
$C_{d\phi\phi}^{V,X}$ & $\phi\bar{\phi}$ & $0.38^{+0.13}_{-0.15}$ & $7.2\pm 1.1$ & $7.1\pm 1.1$ \\ \hline

$C_{d\phi}^{S,X}$ & $\phi$ & $2.1_{ -0.1}^{+0.1}$ & \multicolumn{2}{c}{$(1.8\pm 0.2)\times 10^{-6}$} \\ \hline

$C_{dV}^{V,X}$ & \multirow{2}{*}{$V$} & \multirow{2}{*}{$2.1_{ -0.1}^{+0.1}$} & $(1.4\pm 0.1)\times 10^{-8}$ & $(1.4\pm 0.2)\times 10^{-8}$ \\

$C_{dV}^{T,X}$ & &  & $(5.5\pm 0.5)\times 10^{-3}$ & $(3.5\pm 0.9)\times 10^{-3}$ \\ \hline
\end{tabular}
\caption{Best-fit masses and couplings for the viable scenarios from the global fit to Belle~II, BaBar and ALEPH data.}
\label{tab:best-fit}
\end{table}

For the statistical analysis, we write the total number of expected events as
\begin{align}
n_i & = \mu (1 + \theta_i^{\text{SM}}) s_i^{\text{SM}} + (1 + \theta_i^{X})s_i^{X} \nonumber\\
& \quad + \sum_b \tau_b (1 + \theta_i^{b}) b_i\,,
\end{align}
which adds the contributions of the SM ($s_i^{\text{SM}}$), the light new states ($s_i^X$), and the three background components ($b_i$). The SM signal and backgrounds are multiplied globally by the signal strength $\mu$ and normalisation factor $\tau_b$, respectively. Systematic and Monte-Carlo statistical uncertainties are accounted for by shifting the individual contributions in each bin $i$ by the nuisance parameter $\theta_i^x$ for $x = {\rm SM}, X, b$.

\begin{figure}[t!]
  \centering
  \includegraphics[width=\columnwidth]{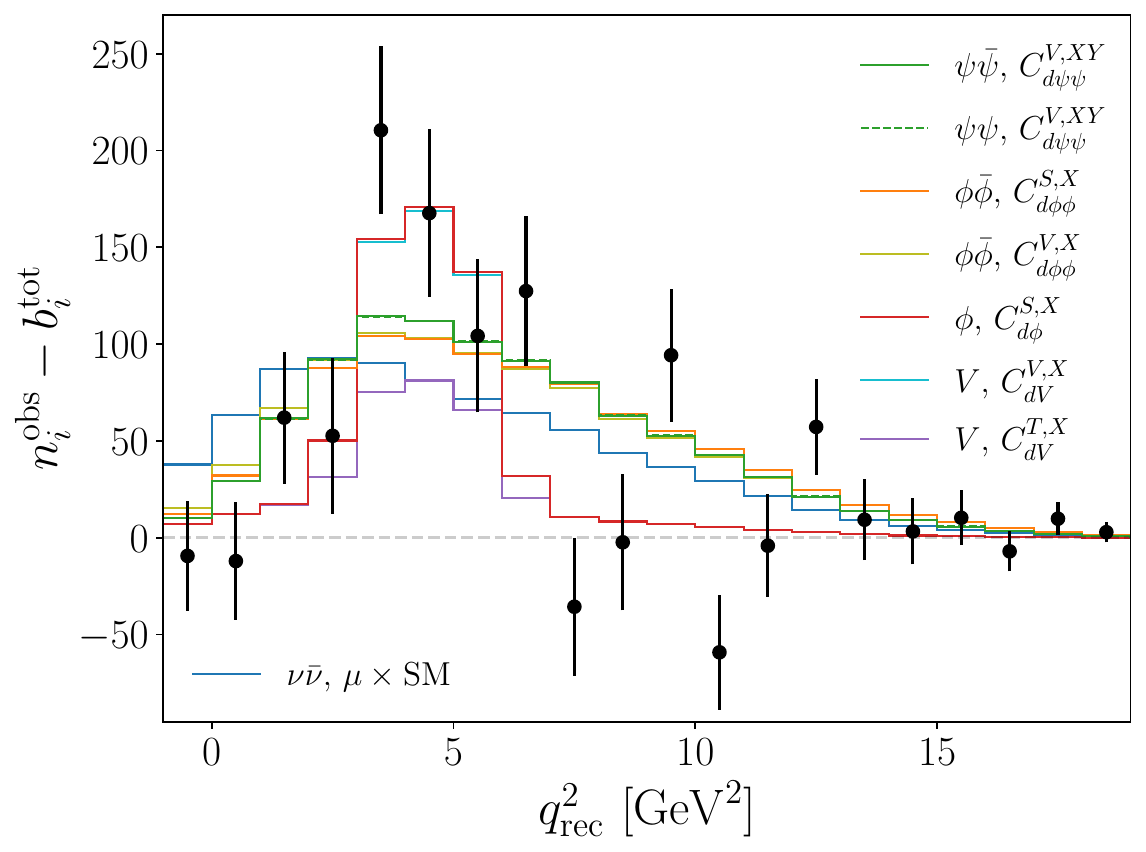}
  \caption{The predicted number of SM + $\sum X$ signal events for $B\to K E_{\text{miss}}$ in bins of $q_{\text{rec}}^{2}$ for the best-fit masses and couplings in Table~\ref{tab:best-fit}, accounting for the Belle~II, BaBar and LEP data. We compare to the number of observed events (black dots) at Belle~II with the backgrounds subtracted.}
  \label{fig:signal_rates}
\end{figure}

For the number of observed events in each bin, $n_i^{\text{obs}}$, we take the results of the ITA in the full signal region, $\eta(\text{BDT}_2) > 0.92$. With this data, we then construct the binned likelihood,
\begin{align}
\label{eq:BelleII-likelihood}
L_{\text{SM}+X}^{\text{Belle~II}} &= \prod_{i,x,b} P\big(n_i^{\text{obs}}, n_i\big)\mc{N}\left(\boldsymbol{\theta}^x; \Sigma^x\right)\mathcal{N}\left(\tau_b; \sigma_b^2\right)\,,
\end{align}
as the product of Poisson probabilities in each bin, a multivariate Gaussian distribution for each signal and background component $x$, and a univariate Gaussian for the overall background normalisation. The covariance matrix of the SM signal, $\Sigma^{\text{SM}}$, is estimated using Monte-Carlo methods and accounts for uncertainties and correlations between bins arising from efficiency and form factor uncertainties. The covariance matrix of each background component, $\Sigma^b$, is found by rescaling $\Sigma^{\text{SM}}$ to reproduce the simulation statistical uncertainties in Fig.~17 of Ref.~\cite{Belle-II:2023esi}. For the covariance matrix for the NP signal, we take for simplicity Poissonian uncertainties and neglect correlations between bins, $(\Sigma^{X})_{ij} = s_i^X\delta_{ij}$. Finally, the systematic uncertainty on the overall background, $\sigma_b$, is chosen to reproduce (in the absence of NP) the value of the profile log-likelihood at $\mu = 0$ in Fig.~16 of Ref.~\cite{Belle-II:2023esi}. As a check of our method, we find a best-fit signal strength of $\mu = 5.3 \pm 1.5$ for the $\mu \times \text{SM}$ hypothesis, in good agreement with Ref.~\cite{Belle-II:2023esi}.

\begin{figure}[t!]
  \centering
  \includegraphics[width=0.9\columnwidth]{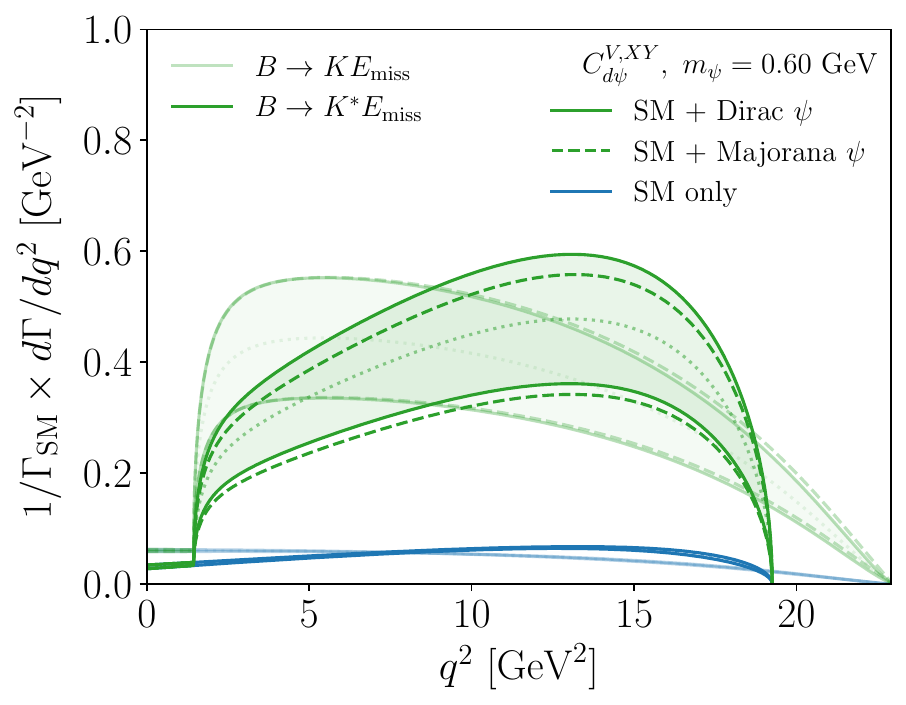}
  \caption{SM and SM + $\psi\bar{\psi}$ (Dirac) or $\psi\psi$ (Majorana) predictions for the differential decay rate of $B\to K^{(*)}E_{\text{miss}}$ divided by the total SM decay rate, given the best-fit mass and couplings from the combined likelihood fit.}
  \label{fig:diff-BRs-fermion}
\end{figure}

With the Belle~II binned likelihood, we now explore the SM + $\sum X$ scenarios involving one invisible final state in Eq.~\eqref{eq:final_states} interacting via one of the chiral basis couplings in Eqs.~\eqref{eq:vector_quark}--\eqref{eq:tensor_quark}. In Fig.~\ref{fig:mass-fit}, we show the result of minimising the (negative) log-likelihood with respect to the chiral basis coupling $C_X$ and nuisance parameters $\boldsymbol{\theta}$ for each mass $m_X$, i.e.
\begin{align}
\label{eq:LL_ratio}
t_X(m_X) = - 2 \ln \frac{L_{\text{SM}+X}^{\text{Belle~II}}(m_X, \hat{C}_X, \hat{\boldsymbol{\theta}})}{L_{\text{SM}}^{\text{Belle~II}}(\hat{\boldsymbol{\theta}})}\,,
\end{align}
where we have normalised with respect to the profiled SM only likelihood with $\mu = 1$. As shown in Fig.~\ref{fig:mass-fit}, all NP scenarios considered here can improve the quality of the fit relative to the $\mu \times \text{SM}$ (blue, dashed) depending on the value of $m_X$. The masses giving the best fit to the Belle~II ITA data for the chiral basis couplings, shown in Table~\ref{tab:best-fit}, are the same as for the parity basis couplings in Ref.~\cite{Bolton:2024egx}. Among the NP scenarios, the two-body decays with $m_{\phi/V} = 2.1$~GeV provide the optimal fit to the data ($4.5\sigma$ with respect to the SM only). It is clear that a $q^2_{\text{rec}}$ peak at $4.4$~$\text{GeV}^2$, with a Gaussian spread due to smearing, is the most efficient at populating the bins containing the Belle~II excess, as illustrated in Fig.~\ref{fig:signal_rates}.

\begin{figure}[t!]
  \centering
  \includegraphics[width=0.9\columnwidth]{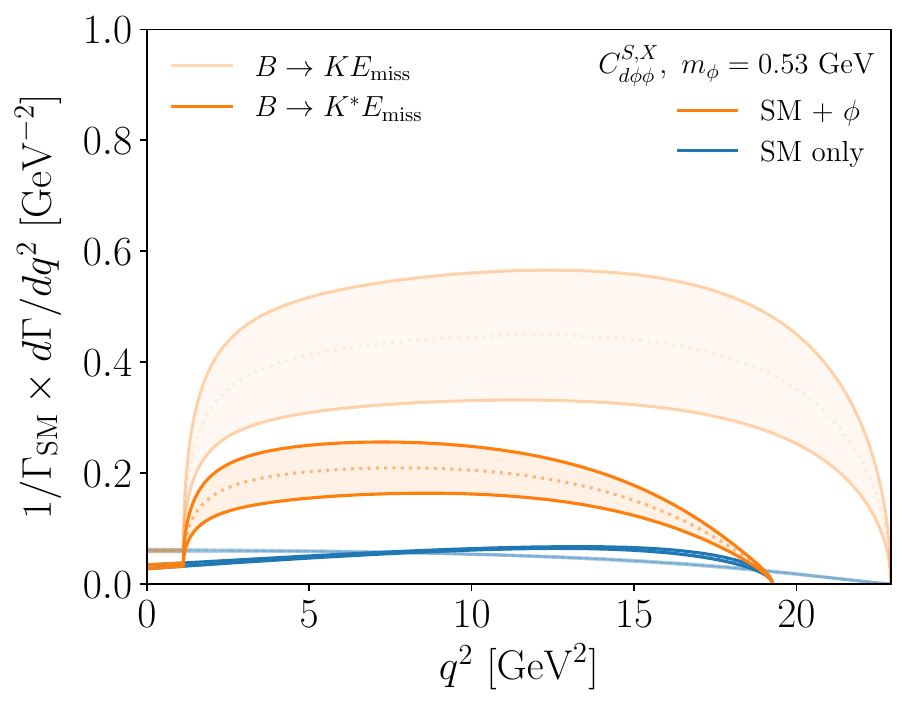}
  \includegraphics[width=0.9\columnwidth]{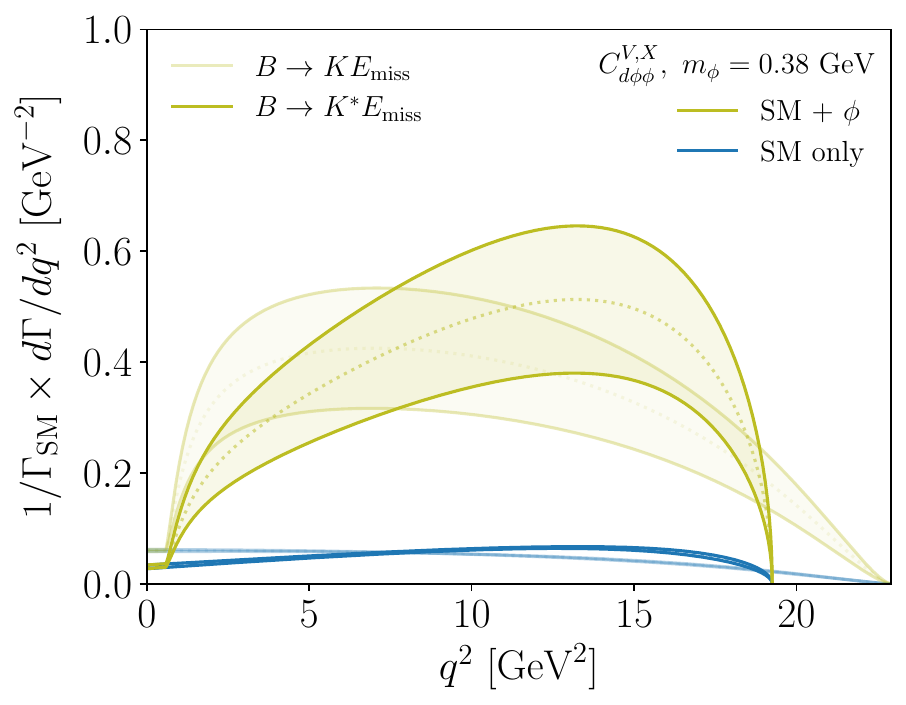}
  \caption{SM and SM + $\phi\bar{\phi}$ predictions for the differential decay rate of $B\to K^{(*)}E_{\text{miss}}$ divided by the total SM decay rate, given the best-fit masses and couplings (scalar and vector couplings above and below, respectively) from the combined likelihood fit.}
  \label{fig:diff-BRs-scalar}
\end{figure}

For the other NP scenarios (in order of decreasing significance), the optimal settings are $m_\psi = 0.60$~GeV for the invisible Dirac or Majorana fermion pair ($3.6\sigma$), $m_\phi = 0.38$~GeV for the invisible complex scalar pair via the vector coupling ($3.4\sigma$) and $m_\phi = 0.53$~GeV for the real/complex scalar pair via the scalar coupling ($3.4\sigma$). For the fermions and scalars produced via Eq.~\eqref{eq:vector_quark}, the differential rates are peaked at $q^2 \sim 4$--$8~\text{GeV}^2$, as seen in Figs.~\ref{fig:signal_rates},~\ref{fig:diff-BRs-fermion} and~\ref{fig:diff-BRs-scalar} (below), respectively. While the tails are significant for $q^2 > 10~\text{GeV}^2$ and could lead to too many events in the high-$q^2$ bins, they are ultimately suppressed by the detection efficiency $\epsilon(q^2)$. For the scalars produced via Eq.~\eqref{eq:scalar_quark}, the peak is less prominent and the whole distribution flatter, as seen in Fig.~\ref{fig:diff-BRs-scalar} (above). The $\epsilon(q^2)$ suppression nevertheless provides a sufficiently peaked distribution in $q^2_{\text{rec}}$, which is made evident in Fig.~\ref{fig:signal_rates}. The differential rates in Figs.~\ref{fig:diff-BRs-fermion} and~\ref{fig:diff-BRs-scalar} will be discussed in more detail in Sec.~\ref{sec:results}.

For the best-fit masses in Table~\ref{tab:best-fit}, the Belle~II ITA data can also be used to determine the best-fit values of the couplings to the invisible sector in Eqs.~\eqref{eq:vector_quark}--\eqref{eq:tensor_quark}. However, for a single non-zero coupling in the chiral basis, $C_X \neq 0$, the decay rate for $B^+ \to K^+ E_{\text{miss}}$ is guaranteed to be correlated with other $b\to s E_{\text{miss}}$ observables. For example, the hadronic matrix elements of the $B\to K$ and $B\to K^*$ transitions are simultaneously non-zero for the $(\bar b \Gamma P_X s)$ quark current with $\Gamma \in \{1, \gamma_\mu, \sigma_{\mu\nu}\}$. Thus, cancellations should occur between different coefficients in the chiral basis to suppress one of the two processes; e.g., the scenario $C_{d\phi}^{V,L} = C_{d\phi}^{V,R}$ (equivalent to $g_S \neq0$ and $g_P = 0$ in the parity basis) giving $\mathcal{B}(B\to K^*\phi) = 0$ for sizable $B\to K\phi$ rates. With only one non-zero chiral basis coupling considered at a time, we cannot explore these directions in the parameter space. For the best-fit values of the couplings we therefore account for the upper bounds on the $B\to K^*E_\text{miss}$ and $B_s \to E_\text{miss}$ processes. It will transpire that only the best-fit value of the dipole coupling $C_{dV}^{T,X}$ is significantly affected by the additional $B\to K^{*}E_\text{miss}$ constraint.

We firstly consider the constraints from $B\to K^* E_{\text{miss}}$. The experimental information that can be extracted for the $B\to K^*(\to K\pi)\sum X$ process is fully described by the double differential decay rate
\begin{align}
\label{eq:double_diff_BtoKs}
\frac{d^2\Gamma}{dq^2 d\cos\theta_K} &= \frac{1}{2}\frac{d\Gamma}{dq^2}\Big(1 + (3F_L - 1)D_{0,0}^2\Big) \,,
\end{align}
where we have integrated over the angular variables of the invisible state(s). Per convention, we define $\theta_K$ as the angle between the $K^*$ direction in the $B$ rest frame and the $K$ direction in the $K^*$ rest frame. The $\cos\theta_K$ dependence is then contained in the Wigner $D$-function, $D_{0,0}^2 = \frac{1}{2}(3\cos^2\theta_K - 1)$. The differential rate $d\Gamma/dq^2$ can be separated into contributions from transversely ($T$) and longitudinally ($L$) polarised intermediate on-shell $K^*$. In principle, only two out of three $K^*$ polarisation states can be disentangled from the $K\pi$ final state, with these two conventionally chosen~\cite{Bobeth:2012vn}. In other words, we have
\begin{align}
\label{eq:BtoKs_quantities}
\frac{d\Gamma}{dq^2} = \frac{d\Gamma_T}{dq^2} + \frac{d\Gamma_L}{dq^2} \,, \quad F_L = \frac{d\Gamma_L}{dq^2}\bigg/\frac{d\Gamma}{dq^2} \,,
\end{align}
where $F_L$ is the longitudinal polarisation fraction of the process. In experiments, the quantities in Eq.~\eqref{eq:BtoKs_quantities} are measured in bins of $q^2$. For example, $F_L$ can be evaluated for the bin $[q^2_i, q^2_{j}]$, with $\Delta q^2 \equiv q_j^2 - q_i^2$, as
\begin{align}
\label{eq:FL_bin}
\braket{F_L}_{\Delta q^2} = \bigg(\int_{q^2_i}^{q^2_{j}} dq^2\frac{d\Gamma_L}{dq^2}\bigg)\bigg/\bigg(\int_{q^2_i}^{q^2_{j}} dq^2\frac{d\Gamma}{dq^2}\bigg)\,.
\end{align}
We can also define $\braket{F_L}$, the longitudinal polarisation fraction for the entire kinematically-allowed $q^2$ range, as Eq.~\eqref{eq:FL_bin} with $q^2_i = 0$ and $q^2_{j} = (m_B - m_{K^*})^2$.

In the SM, $B \to K^{*}\nu\bar{\nu}$ has the rate
\begin{align}
\label{eq:BtoKsSM}
\frac{d\Gamma(B \to K^* \nu\bar{\nu})_{\text{SM}}}{dq^2} = \frac{|\vec{p}_{K^*}| q^2}{64 \pi^3} \big|C_{d\nu}^{V,LL}\big|^2 \big(f_{T}^2 + f_{L}^2\big)\,,
\end{align}
where the form factor dependence has been separated into transverse and longitudinal parts, respectively, as
\begin{align}
f_{T}^2 & \equiv \frac{2 |\vec{p}_{K^*}|^2}{(m_B + m_{K^*})^2}V^{2} + \frac{(m_B + m_{K^*})^2}{2m_B^2} A_1^{2} \,, \label{eq:fT}\\
f_{L}^2 & \equiv \frac{16m_{K^*}^2}{q^2} A_{12}^{2} \,,
\label{eq:fL}
\end{align}
with the vector ($V$) and axial-vector ($A_1$, $A_{12}$) $B \to K^*$ form factors. With the input parameters below Eq.~\eqref{eq:CdvLL} and the form factors determined via the BSZ fit results of~\cite{Gubernari:2023puw}, the SM branching fraction for the neutral mode is found to be
\begin{align}
\label{eq:B_BtoKs_SM}
\mathcal{B}(B^0 \to K^{*0} \nu\bar{\nu})_{\text{SM}} &= (8.95 \pm 0.89 \pm \delta)\times 10^{-6} \,,
\end{align}
using $\tau_{B^0} = 1.519(4)$~ps. The hadronic and parametric uncertainties are separated as before, with $\delta = 0.45$. The SM expression for the longitudinal polarisation fraction, $F_L^{\text{SM}}$, can be obtained from Eq.~\eqref{eq:BtoKsSM}. Without integrating the numerator and denominator over $q^2$, the common prefractor of the $f_T^2$ and $f_L^2$ terms cancels, resulting in the expression $F_{L}^{\text{SM}} = f_L^2/(f_L^2 + f_T^2)$. For the limiting $q^2$ values, $F_L^{\text{SM}}(0) = 1$ and $F_L^{\text{SM}}(q_{\text{max}}^2) = 0.308 \pm 0.045$. Keeping the full $q^2$ dependence of the numerator and denominator in Eq.~\eqref{eq:BtoKs_quantities} before integrating over the full kinematically allowed $q^2$ region, we obtain
\begin{align}
\label{eq:FL_SM_val}
\braket{F_L}_{\text{SM}} &= 0.466 \pm 0.024 \,.
\end{align}
Only the form factor uncertainties contribute in Eq.~\eqref{eq:FL_SM_val}, as the parametric uncertainties cancel. 

The NP interactions in Eqs.~\eqref{eq:vector_quark}--\eqref{eq:tensor_quark} also induce the process $B\to K^* \sum X$, with the (differential) decay rates for transversely and longitudinally polarised $K^*$ given in App.~\ref{app:decays}. The NP contributions depend on the additional axial-vector ($A_0$) and tensor ($T_1$, $T_2$ and $T_{23}$) $B\to K^*$ form factors. With these predictions, we can now implement the constraints from experimental upper bounds on the $B\to K^*\nu\bar{\nu}$ branching fraction. 

The most stringent current upper bound, quoted in Eq.~\eqref{eq:Belle-result}, is from the Belle experiment~\cite{Belle:2017oht}. Unfortunately, this analysis did not provide binned distributions in $q^2$, preventing a shape analysis for the SM + $\sum X$ scenarios. We therefore consider the results of the complementary BaBar analysis~\cite{BaBar:2013npw} which, despite setting weaker upper limits on the $B^\pm$ and $B^0$ decay branching fractions, provided $s_B \equiv q^2$ distributions. To identify the decays $K^{*+}\to K^+\pi^0/K_S^0\pi^+$ and $K^{*0}\to K^+\pi^-/K_S^0
\pi^0$, the analysis used the hadronic tagging method. However, for the reasons outlined in App.~D of Ref.~\cite{Bolton:2024egx}, we consider only the $B^0$ decay mode with $K^{*0}$ decaying dominantly to $K^+\pi^-$, neglecting smearing effects. The backgrounds and their $q^2$-dependence are extracted from Fig.~5, while the efficiencies $\epsilon(q^2)$ are taken from Fig.~6. We compute the uncertainties and correlations between bins similarly to the Belle~II analysis in order to reproduce the SM upper limit in Table~IV. Now, we can calculate the expected number of events in the SM + $\sum X$ scenario as input to a binned likelihood, analogous to Eq.~\eqref{eq:BelleII-likelihood}.

\begin{figure}[t!]
  \centering
  \includegraphics[width=0.9\columnwidth]{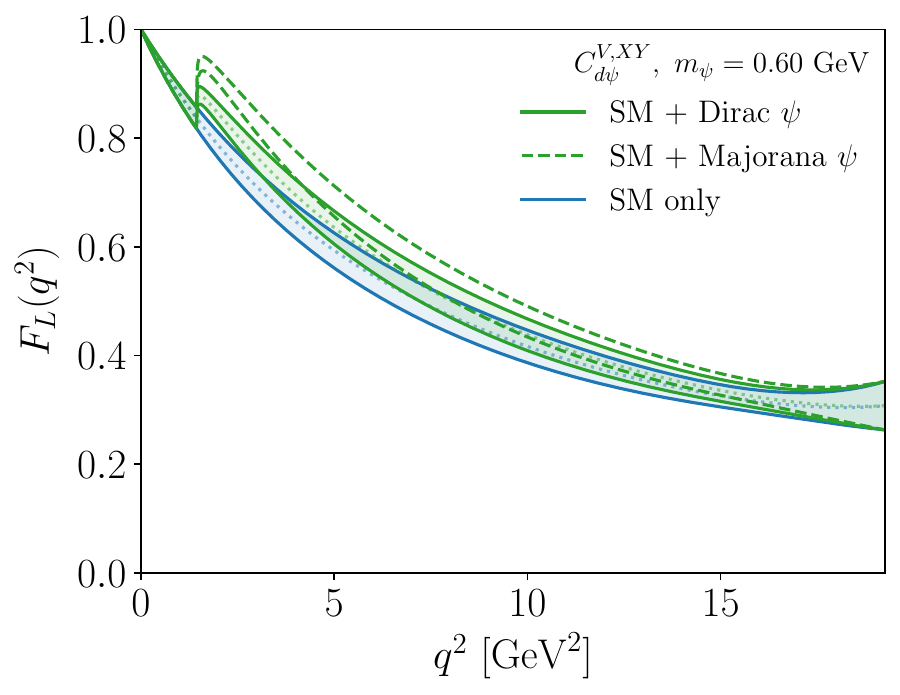}
  \caption{SM and SM + $\psi\bar{\psi}$ (Dirac) or $\psi\psi$ (Majorana) predictions for the $K^*$ longitudinal polarisation fraction $F_L$, given the best-fit mass and couplings as in Table~\ref{tab:best-fit}.}
  \label{fig:FL_fermion}
\end{figure}
\begin{figure}[t!]
  \centering
  \includegraphics[width=0.9\columnwidth]{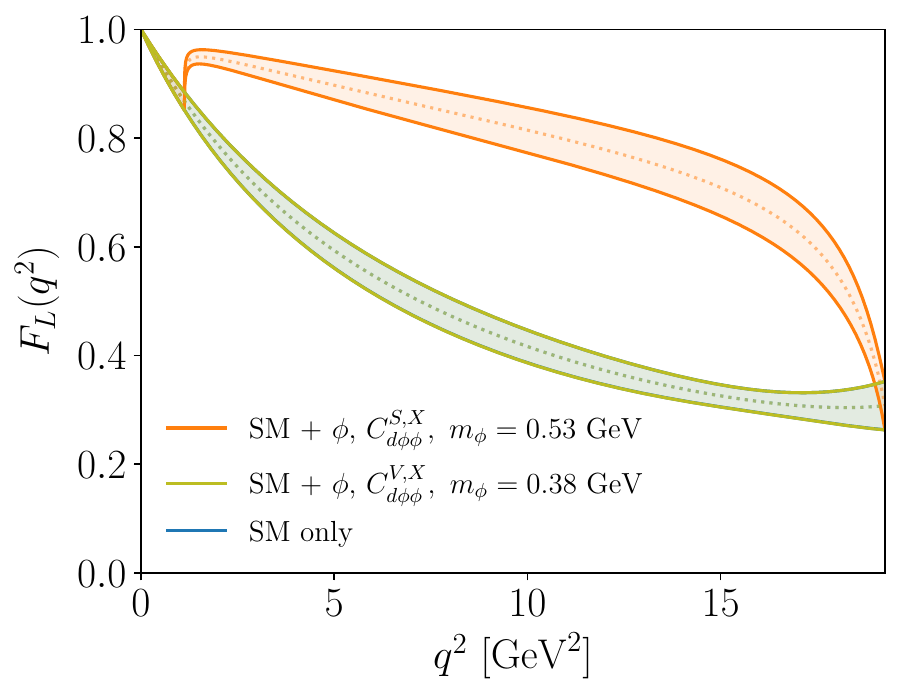}
  \caption{SM and SM + $\phi\bar{\phi}$ predictions for the $K^*$ longitudinal polarisation fraction $F_L$, given the best-fit mass and couplings as in Table~\ref{tab:best-fit}. The predictions for the SM only and $C_{d\phi\phi}^{V,X} \neq 0$ scenarios overlap.}
  \label{fig:FL_scalar}
\end{figure}

Finally, we implement the constraints on the invisible decay $B_s \to E_{\text{miss}}$. For the three-body $B^+ \to K^+ \sum X$ decays induced via Eqs.~\eqref{eq:vector_quark}--\eqref{eq:tensor_quark}, the process $B_s\to\sum X$ is also possible. In the SM, the decay $B_s \to \nu\bar{\nu}$ is helicity suppressed and therefore vanishes for strictly massless neutrinos. For the light neutrino masses inferred from oscillation data, it is still extremely suppressed, and in fact the $B_s \to \nu\bar{\nu}\nu\bar{\nu}$ process is larger~\cite{Bhattacharya:2018msv}, albeit still well below experimental sensitivities. The decay mode is therefore an interesting probe of new physics. The $B_s \to \sum X$ decay rates for the NP scenarios $\sum X \in \{\phi\phi,\phi\bar{\phi},\psi\psi,\psi\bar{\psi}\}$ are given in App.~\ref{app:decays} and are non-vanishing for non-zero chiral basis couplings $C_{d\phi\phi}^{S,X}$, $C_{d\phi\phi}^{V,X}$ and $C_{d\psi}^{V,XY}$. In Ref.~\cite{Alonso-Alvarez:2023mgc}, the ALEPH inclusive search for $b\to \tau^{-}\bar{\nu}_\tau X$ at LEP~\cite{ALEPH:2000vvi} was recast to obtain the upper limit $\mathcal{B}(B_s \to E_{\rm miss}) < 5.6 \times 10^{-4}$ (90\% CL). We construct a simple Gaussian log-likelihood to also enforce this constraint on the chiral basis couplings.

We now explore the best-fit couplings implied by the data for the NP scenarios. In each case, the mass is fixed to the central value in Table~\ref{tab:best-fit} and the scale of NP is set to $\Lambda = 10$~TeV, and we construct a global likelihood incorporating the Belle~II, BaBar and ALEPH data. In the rightmost column of Table~\ref{tab:best-fit}, we show the best-fit coupling values for the Belle~II data only (left) and with the BaBar and ALEPH data added (right), found at the global minimum of the likelihood. In the cases where the addition of the BaBar and ALEPH data does not change the best-fit value at this level of accuracy, we show a common value. It can be seen that, for most of the scenarios, the inclusion of the BaBar and ALEPH data only leads to a marginal decrease in the best-fit value. However, for the dipole-like coupling $C_{dV}^{T,X}$, a large reduction is seen, because the best-fit coupling value for the Belle~II excess is in tension with the BaBar data.

For the best-fit masses and couplings (including the Belle~II, BaBar and LEP data in the fit), the predicted $B\to K E_{\text{miss}}$ signal rates (including the SM + $\sum X$ contributions) in bins of $q_{\text{rec}}^{2}$ at Belle II are shown in Fig.~\ref{fig:signal_rates} for each NP scenario. These are compared to the number of observed events (black dots) at Belle~II with the backgrounds subtracted. We have already noted that the best-fit masses are strongly influenced by the excess of events at $3< q^2_{\text{rec}}/\text{GeV}^2 < 5$. The best-fit couplings are in turn determined by the size of the excess. If only the Belle~II data is included in the fit, the signal predictions in the two-body scenarios would be identical. However, with the inclusion of the BaBar data, the signal prediction for the dipole-like coupling $C_{dV}^{T,X}$ is smaller with respect to the other two-body scenarios.

\begin{figure}[t!]
  \centering
  \includegraphics[width=0.9\columnwidth]{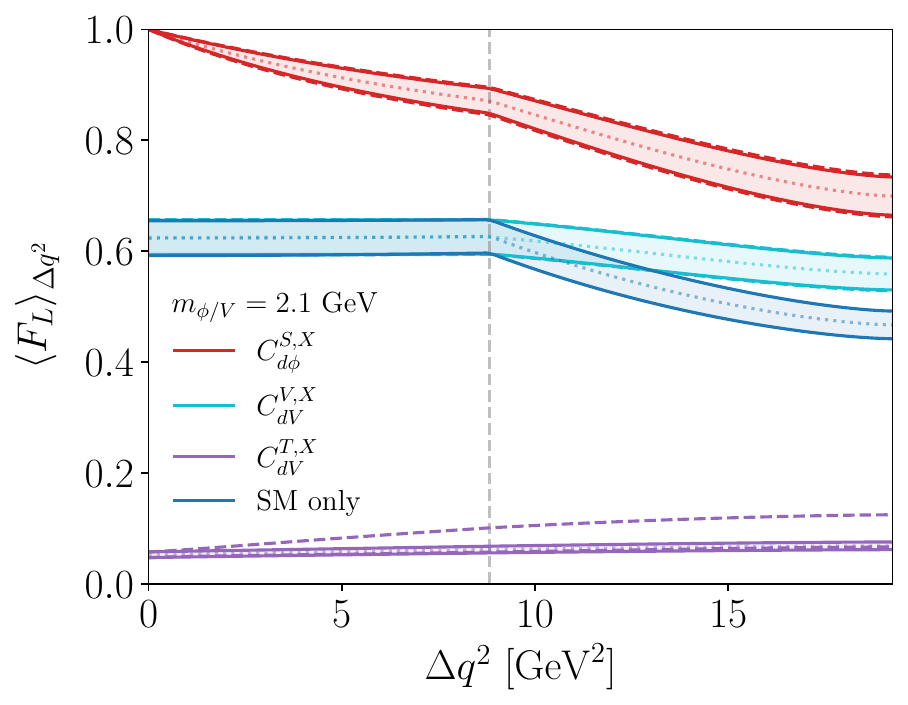}
  \caption{SM + $\phi/V$ predictions for the binned $K^*$ longitudinal polarisation fraction $\langle F_L \rangle_{\Delta q^2}$ as a function of the bin size for a bin around $q^2=m_{\phi/V}^2=(2.1~\mathrm{GeV})^2$ , given the best-fit mass and couplings as in Table~\ref{tab:best-fit}. The gray dashed line indicates the bin size ($\Delta q^2=2m_{\phi/V}^2$) at which the bin reaches $q^2 = 0~\mathrm{GeV}^2$, changing from a symmetrical to an asymmetrical bin with respect to $q^2=m_{\phi/V}^2$.}
  \label{fig:FL_2body}
\end{figure}
%

%%%%%%%%%%%%%%%%%%%%%%%%%%%%%%%%%%%%%%%%%%
%
\section{New Physics Predictions for $B \to K^* E_{\text{miss}}$ Observables}
\label{sec:results}
%
%%%%%%%%%%%%%%%%%%%%%%%%%%%%%%%%%%%%%%%%%%

%
\begin{figure*}[t!]
  \centering
  \includegraphics[width=\linewidth]{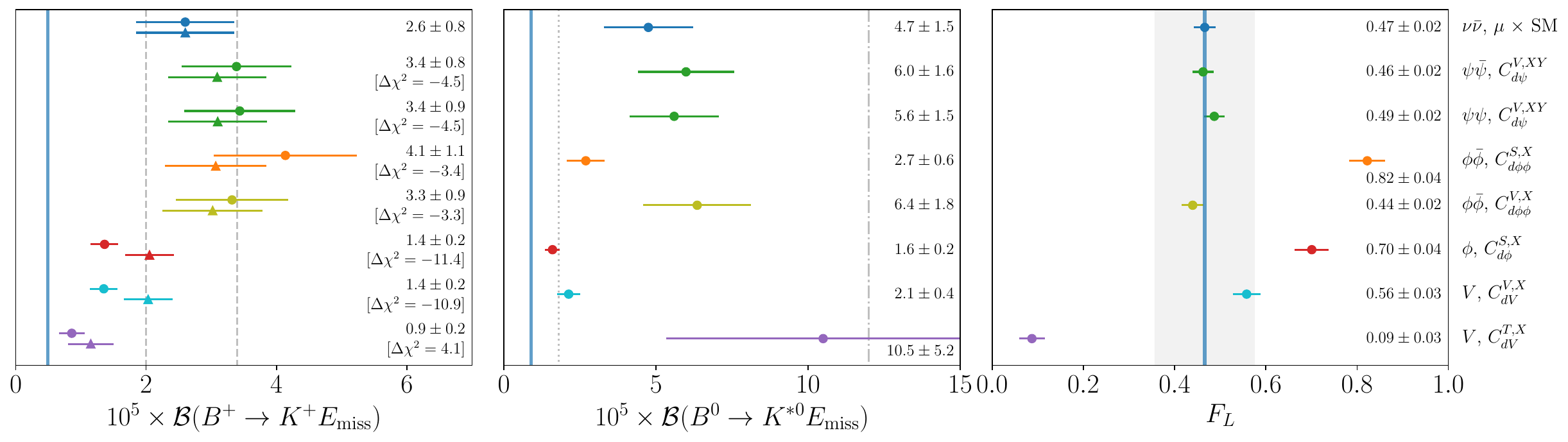}
  \caption{SM and SM + $\sum X$ predictions for the $B^+\to K^+ E_{\text{miss}}$ (left) and $B^0\to K^{*0} E_{\text{miss}}$ (centre) branching fractions and $B^0\to K^{*0} E_{\text{miss}}$ longitudinal polarisation fraction (right). Belle~II constraints~\cite{Belle-II:2023esi} are indicated by the dashed gray lines (left), the BaBar~\cite{BaBar:2013npw} and Belle~\cite{Belle:2017oht} upper limits are indicated by the dot-dashed and dotted lines respectively (centre) and the expected sensitivity~\cite{Belle-II:2018jsg} of Belle~II for the longitudinal polarisation fraction is indicated by the light gray shaded region (right).}
  \label{fig:BR_FL_summary}
\end{figure*}

In this section, we now examine the implications for $B\to K^*E_{\text{miss}}$ observables given the best-fit masses and couplings in Table~\ref{tab:best-fit}. Of the $b \to s$ transition processes, $B\to K^*E_{\text{miss}}$ will be measured to a high level of accuracy the soonest by the Belle~II experiment~\cite{Belle-II:2018jsg}, and therefore will be able to test (i.e., exclude or confirm) the NP scenarios implied by the $B^+\to K^+E_{\text{miss}}$ excess. The prospects for improved bounds on $B_s \to E_{\text{miss}}$ are longer term, with measurements by Belle~II at the $\Upsilon(5S)$ resonance~\cite{Belle-II:2018jsg} and opportunities during the Tera-$Z$ run of FCC-ee~\cite{Amhis:2023mpj} and/or CEPC~\cite{Li:2022tov,CEPCPhysicsStudyGroup:2022uwl}. It is expected that the first bounds on $\Lambda_b \to \Lambda E_{\text{miss}}$ will also be obtained only at these future $Z$-factories. With this longer timeline in mind, we do not examine further the expected signals of $B_s \to E_{\text{miss}}$ and $\Lambda_b \to \Lambda E_{\text{miss}}$ in this work.

Firstly, in Figs.~\ref{fig:diff-BRs-fermion} and~\ref{fig:diff-BRs-scalar}, we plot the differential rates for the three-body decay contributions to $B\to KE_{\text{miss}}$ (light shaded) and $B\to K^*E_{\text{miss}}$ (dark shaded) for the best-fit masses and couplings in Table~\ref{tab:best-fit} (with the values from the Belle~II + BaBar + LEP fit). In each plot, we show the SM only and SM + $\sum X$ predictions, normalised to the total SM rate $\Gamma_{\text{SM}}$. The bands indicate the $\pm 1\sigma$ uncertainties from the coupling and form factor uncertainties (the former dominating for the SM + $\sum X$ rates). It is clear that to adequately explain the Belle~II data, the SM + $\sum X$ predictions for $B^+\to K^+ E_{\text{miss}}$ must be considerably above the SM only rates. Given the correlation with $B\to K^{*} E_{\text{miss}}$, these rates also see a significant increase with respect to the SM. This is particularly the case for $B\to K^*\psi\psi/\psi\bar{\psi}$ and $B\to K^*\phi\bar{\phi}$ induced via the coupling $C_{d\phi\phi}^{V,X}$. While the distributions peak at high $q^2$, which could see significant reductions due to the detection efficiency, a large rate is still seen above the production threshold $q^2 = 4m_X^2$. A small but non-negligible difference is seen between the Dirac (solid) and Majorana (dashed) fermion predictions for $B\to K^*E_{\text{miss}}$. For $B\to K^*\phi\phi/\phi\bar{\phi}$ induced via the coupling $C_{d\phi\phi}^{S,X}$, the $B \to K^*E_{\text{miss}}$ is relatively smaller compared to the other three-body decay scenarios.

Next, in Figs.~\ref{fig:FL_fermion}--\ref{fig:FL_scalar}, we show the predicted values of the longitudinal polarisation fraction $F_L(q^2)$ for the best-fit couplings and masses of the three-body decay scenarios. In Fig.~\ref{fig:FL_fermion}, $F_L$ values for Dirac (solid, green) and Majorana (dashed, green) fermions in the final state are compared to the SM expectation (blue). Above the kinematic threshold, $F_L$ deviates (more prominently in the Majorana case) from the SM prediction, before tending back towards the limiting SM value at the kinematic endpoint. To understand this behaviour, Eqs.~\eqref{eq:BKs_psipsi_T} and~\eqref{eq:BKs_psipsi_L} can be used to find that $F_L$ is schematically of the form
\begin{align}
\label{eq:FL_threebody}
F_L = \frac{f_L^2 + f_{X}^2}{f_L^2 + f_T^2 + f_X^2}\geq F_L^{\text{SM}}\,,
\end{align}
where $f_X^2 \propto A_0^2$ vanishes at the boundaries of the kinematically allowed region for $B\to K^* \psi\psi/\psi\bar{\psi}$. In Fig.~\ref{fig:FL_scalar}, we instead plot the $F_L$ values for the scalar three-body decay scenarios. For the $B\to K^*\phi\phi/\phi\bar{\phi}$ process induced by the coupling $C_{d\phi\phi}^{S,X}$, the transverse rate in Eq.~\eqref{eq:BKs_phiphi_T} vanishes and the longitudinal rate in Eq.~\eqref{eq:BKs_phiphi_L} is proportional to $A_0^2$. The resulting $F_L$ is then of the same form as Eq.~\eqref{eq:FL_threebody}, with $f_X^2$ leading to an even larger increase of $F_L$ with respect to the SM. Finally, for the $B\to K^*\phi\bar{\phi}$ decay induced by the coupling $C_{d\phi\phi}^{V,X}$, $f_X^2 = 0$ and $F_L$ is identical to the SM. We would like to emphasise, however, that cancellations such as in Eq.~\eqref{eq:FL_threebody} cannot be made for the integrated longitudinal polarisation fraction, $\braket{F_L}$. With the complete $q^2$ dependence of the numerator and denominator integrated over, the value of $\braket{F_L}$ can be different to what is naively expected from Fig.~\ref{fig:FL_fermion}--\ref{fig:FL_scalar}. We will return to this in the discussion of Fig.~\ref{fig:BR_FL_summary}.

In Fig.~\ref{fig:FL_2body}, we study the longitudinal polarisation for the two-body decay scenarios. With $q^2$ fixed to the mass of the scalar or vector boson for the decay $B \to K^* \phi/V$, we consider $\braket{F_L}_{\Delta q^2}$ obtained by integrating the numerator and denominator of $F_L$ over a bin $[q^2_i, q^2_j]$ centred on $q^2 = m_X^2$. In Fig.~\ref{fig:FL_2body}, we show $\braket{F_L}_{\Delta q^2}$ as a function of the bin width $\Delta q^2$, which is increased from $\Delta q^2 = 0$ to the full kinematically-allowed region. The vertical gray dashed line indicates where the lower edge of the bin reaches $q^2_i = 0$, changing from a symmetric to asymmetric bin with respect to $q^2 = m_X^2$. In the limit $\Delta q^2 \to 0$, the SM contributions in the numerator and denominator are infinitesimally small with respect to the two-body decay contributions, which have $\delta$ function support. Using Eqs.~\eqref{eq:BKsphiL} and \eqref{eq:BKsphiT}, the value at $\Delta q^2 = 0$ for $B\to K^*\phi$ is therefore $\braket{F_L}_{\Delta q^2} = 1$. With Eqs.~\eqref{eq:BKsVL} and \eqref{eq:BKsVT}, the values at $\Delta q^2 = 0$ for the $B\to K^*V$ scenario are $\braket{F_L}_{\Delta q^2} = f_L^2/(f_T^2 + f_L^2) = 0.625 \pm 0.032$ and $\braket{F_L}_{\Delta q^2} = 0.052 \pm 0.005$ (with the form factors evaluated at $q^2 = m_V^2$) for $C_{dV}^{V,X}$ and $C_{dV}^{T,X}$, respectively. We note that the coefficients cancel in the ratio, so only the form factor uncertainties contribute. As $\Delta q^2$ increases, the SM contribution in both the numerator and denominator grows, pulling the values of $\braket{F_L}_{\Delta q^2}$ towards the SM value in Eq.~\eqref{eq:FL_SM_val}. For each NP scenario, we show the results for the best-fit couplings taken from the Belle~II only (solid) and Belle~II + BaBar + LEP fits (dashed). The difference in these results clearly only has an impact for $C_{dV}^{T,X}$. From Fig.~\ref{fig:FL_2body}, it is clear that for the $B\to K^*\phi$ and $B\to K^*V$ (induced via $C_{dV}^{T,X}$) scenarios, measuring $\braket{F_L}_{\Delta q^2}$ in narrow bins around $q^2 = m_X^2$ provides more of a smoking-gun deviation from the SM. The opposite is true for $B\to K^*V$ induced via $C_{dV}^{V,X}$, where the whole kinematic range provides more discrimination.

In Fig.~\ref{fig:BR_FL_summary}, we summarise the results discussed so far. For each SM + $\sum X$ scenario with the best-fit masses and couplings in Table~\ref{tab:best-fit} (Belle~II + BaBar + LEP), we depict as circular points the predicted values of the $B^+ \to K^+E_{\text{miss}}$ and $B^{0} \to K^{*0}E_{\text{miss}}$ branching fractions and the integrated longitudinal polarisation fraction $\braket{F_L}$. The $\pm 1\sigma$ error bars incorporate the uncertainties in the best-fit couplings and form factors. To illustrate how these deviate from the SM, we show as blue vertical lines the predicted SM values in Eqs.~\eqref{eq:B_BtoK_SM},~\eqref{eq:B_BtoKs_SM} and~\eqref{eq:FL_SM_val}, respectively. In the leftmost plot, the dashed vertical lines show the range of branching fractions implied by the Belle~II ITA result, $\mathcal{B}(B^+ \to K^+ E_{\text{miss}}) = (2.7 \pm 0.7)\times 10^{-5}$. Our $\mu \times \text{SM}$ central value and uncertainties are in agreement with this band, as expected. However, large deviations for the NP scenarios can be seen, highlighting the importance of the $q^2$ distribution shape analysis in the inference of the mass and coupling of the light invisible field; in other words, a naive implementation of the measured Belle~II branching fraction would have led to incorrect conclusions. To demonstrate this point even further, we plot as triangular points the values with efficiency and smearing effects included (i.e. integrating the differential rate with $f_{q^2_{\text{rec}}}(q^2)$ and $\epsilon(q^2)$ factors) and rescaling by the ratio of the $\mu\times \text{SM}$ branching fraction with and without 
the efficiency and smearing included. The SM + $\sum X$ predictions are still different from the $\mu \times \text{SM}$ scenario and from each other.

\begin{figure*}[t!]
  \includegraphics[width=0.496\linewidth]{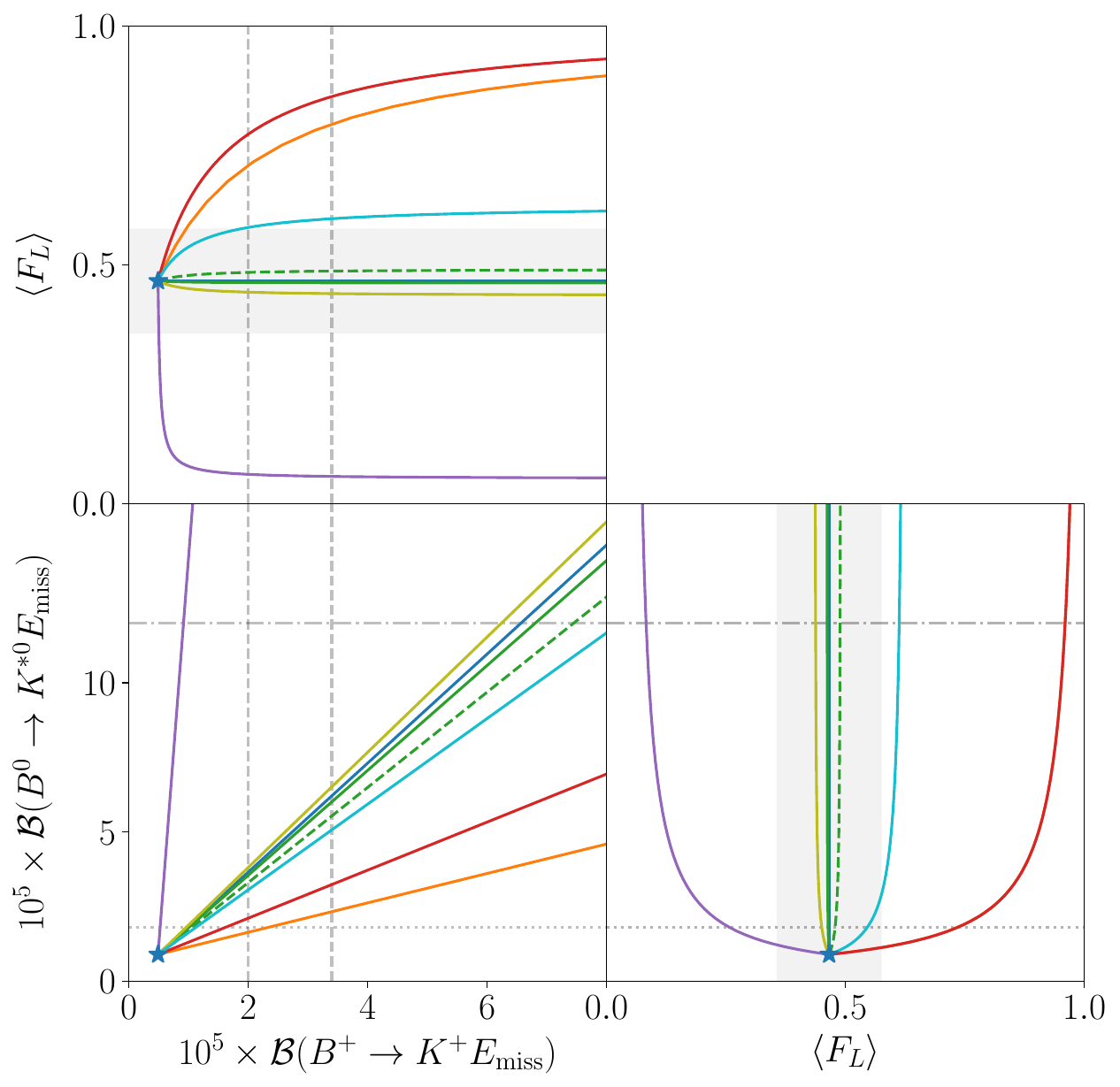}
  \includegraphics[width=0.496\linewidth]{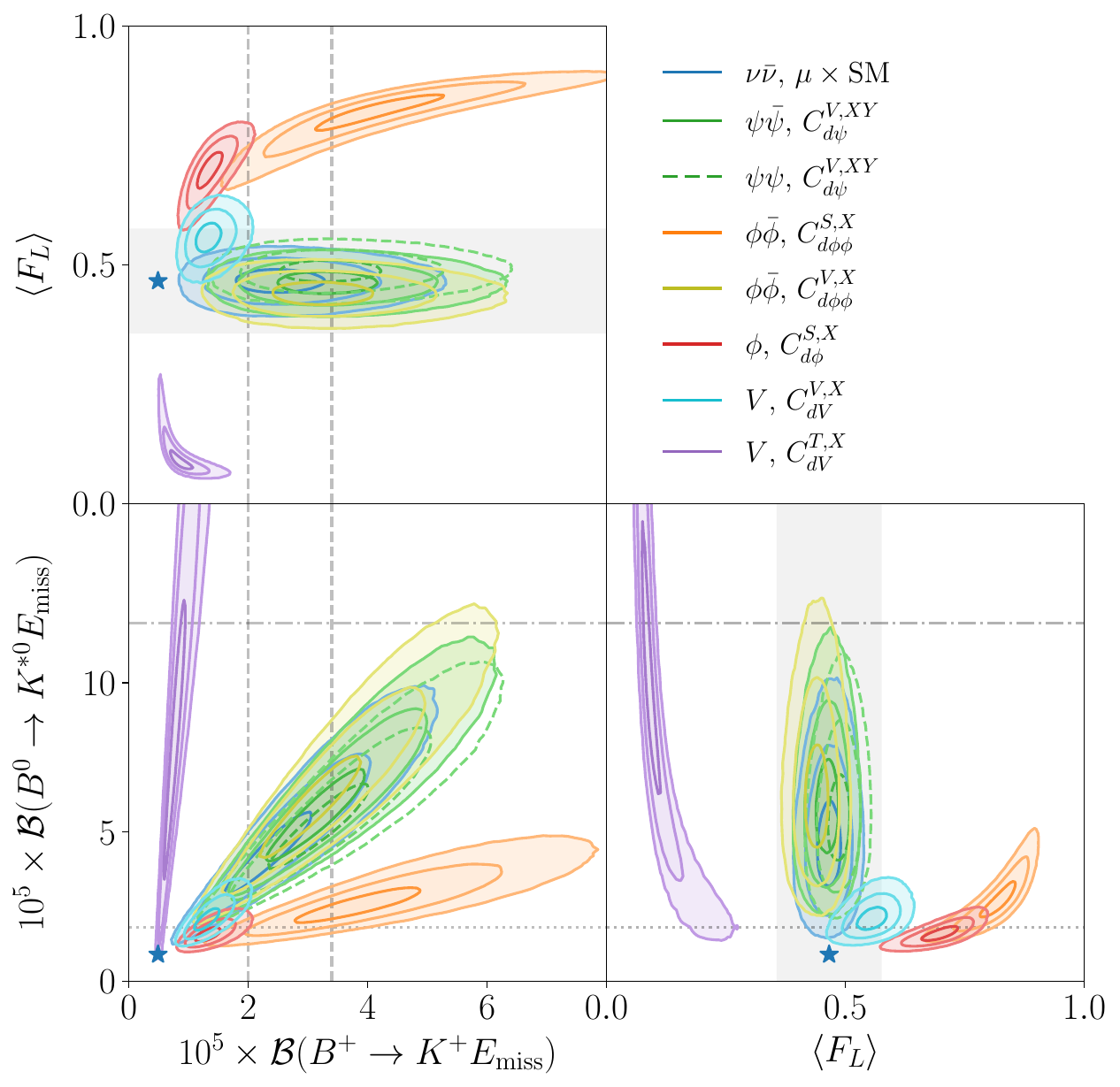}
  \caption{(Left) Correlations between the $B^+\to K^+ E_{\text{miss}}$ and $B^0\to K^{*0} E_{\text{miss}}$ branching fractions and $B^0\to K^{*0} E_{\text{miss}}$ longitudinal polarisation fraction in the SM + $\sum X$ scenarios, with form factor uncertainties omitted, shown with respect to the SM predictions (stars). (Right) $1\sigma$, $2\sigma$ and $3\sigma$ confidence regions for the SM + $\sum X$ scenarios given the best-fit masses and couplings in Table~\ref{tab:best-fit} (Belle~II + BaBar + LEP). The BaBar, Belle and Belle~II constraints/sensitivities match Fig.~\ref{fig:BR_FL_summary}.}
  \label{fig:cornerplot}
\end{figure*}

Turning to the $B\to K^* E_{\text{miss}}$ observables, the largest branching fractions $\mathcal{B}(B^0 \to K^{*0} E_{\text{miss}})$ are seen for the $\sum X \in \{V, \psi\psi,\psi\bar{\psi},\phi\bar{\phi}\}$ final states, with $V$ and $\phi\bar{\phi}$ coupled via $C_{dV}^{T,X}$ and $C_{d\phi\phi}^{V,X}$, respectively. The $\braket{F_L}$ values for these three-body scenarios, however, all lie close to the SM prediction and within the $\pm 0.11$ precision expected by Belle~II~\cite{Belle-II:2018jsg}, indicated as a vertical shaded band. For the other NP scenarios, the branching fraction $\mathcal{B}(B^0 \to K^{*0} E_{\text{miss}})$ is smaller, while larger deviations from the SM value of $\braket{F_L}$ are seen. The $B\to K^*V$ scenario with $C_{dV}^{T,X} \neq 0$ therefore implies large deviations from the SM for both $\mathcal{B}(B^0 \to K^{*0} E_{\text{miss}})$ and $\braket{F_L}$. However, as the BaBar data is in slight tension with the Belle~II excess for this model, the goodness of fit is not competitive with the other NP scenarios (even the $\mu \times \text{SM}$). In the plot, we show the $\Delta \chi^2$, or difference between the minima of the $\mu 
\times \text{SM}$ and SM + $\sum X$ global likelihoods, to indicate the quality of the fit. The two-body decay scenario with a final-state scalar provides the best overall fit.

Finally, it is interesting to explore the correlations between the $B\to KE_\text{miss}$ and $B\to K^* E_\text{miss}$ observables. In Fig.~\ref{fig:cornerplot} (left), we plot the relationships between the $B^+\to K^+ E_{\text{miss}}$ and $B^0\to K^{*0} E_{\text{miss}}$ branching fractions and $B^0\to K^{*0} E_{\text{miss}}$ integrated longitudinal polarisation fraction in the SM + $\sum X$ scenarios, omitting form factor uncertainties. In Fig.~\ref{fig:cornerplot} (right), we complement this by showing the $1\sigma$, $2\sigma$ and $3\sigma$ confidence regions in the same three parameter spaces, comparing to the SM predictions (blue stars). To obtain these regions, as in Fig.~\ref{fig:BR_FL_summary}, we keep the masses fixed to the central values in Table~\ref{tab:best-fit} and incorporate the uncertainties of the best-fit couplings (Belle~II + BaBar + LEP) and form factors. As expected, the branching $B \to KE_{\text{miss}}$ and $B \to K^* E_{\text{miss}}$ are positively correlated in all cases, with gradients ranging from the $\phi\bar{\phi}$ scenario with $C_{d\phi\phi}^{S,X}$ (large $\mathcal{B}(B\to K\phi\bar{\phi})$ and small $\mathcal{B}(B\to K^*\phi\bar{\phi})$) and the $V$ scenario with $C_{dV}^{T,X}$ (vice versa). On the other hand, $\braket{F_L}$ behaves very differently, exhibiting almost no correlation with the two decay rates for $\psi\bar{\psi}$ scenarios as well for the $\phi\bar{\phi}$ scenario with $C_{d\phi\phi}^{V,X}$, a negative correlation for the $C_{dV}^{T,X}$ scenario, and positive correlations for the other three considered cases.

%%%%%%%%%%%%%%%%%%%%%%%%%%%%%%%%%%%%%%%%%%
%
\section{Conclusions}
\label{sec:conclusions}
%
%%%%%%%%%%%%%%%%%%%%%%%%%%%%%%%%%%%%%%%%%%

In this work, we have revisited the implications of a potential excess in the rare decay $B \to K E_\text{miss}$, as indicated by the recent Belle~II measurement, within a general new physics (NP) framework allowing for light invisible final states. Building upon the EFT setup of Ref.~\cite{Kamenik:2011vy}, we have considered the possible scalar, fermion, and vector final state scenarios that can enhance the total rate and, crucially, modify the differential distributions for both $B \to K E_\text{miss}$ and $B \to K^*E_\text{miss}$. 

Our global fits to the Belle~II data, supplemented by constraints from BaBar and LEP, reveal several viable NP scenarios that can yield a better agreement with the measured $B \to K E_\text{miss}$ distribution than the (rescaled) SM. These include two-body final states with a new scalar or vector particle, as well as three-body final states with pairs of (pseudo)scalar or (Majorana/Dirac) fermions. 

We reiterate the conclusions of Ref.~\cite{Bolton:2024egx} that the measured decay spectrum is a key discriminator between different NP interpretations, highlighting that a simple rescaling of the Standard Model (SM) rate ($\mu\times \text{SM}$) is insufficient once the reconstructed distributions in momentum transfer are properly taken into account.

In each case, existing experimental constraints on $B \to K^*E_\text{miss}$ and $B_s \to E_\text{miss}$ can already restrict portions of parameter space, although apart from dipole-like couplings to a spin-1 state, their impact on the best-fit points is limited at present.  

We demonstrate however, that future measurements of $B \to K^*E_\text{miss}$ observables will offer much stronger discrimination among the candidate scenarios. In particular, the total branching fraction and the differential distribution of the longitudinal polarisation fraction $F_L$ can exhibit sizable deviations from the SM predictions in the relevant regions of NP parameter space. 

From a broader perspective, our study highlights, that rare $b$-hadron decays with missing energy remain an excellent probe of physics beyond the SM. New states lighter than a few GeV can produce distinct kinematic imprints that will become experimentally accessible with dedicated searches and improved tagging techniques at Belle~II and currently planned future experimental facilities such as the FCC-ee and CEPC.  
The continued effort to measure (differential) branching fractions and angular observables in $B \to K^{(*)}E_\text{miss}$, as well as to constrain other related processes such as $B_s \to \phi E_\text{miss}$ and $\Lambda_b \to \Lambda E_\text{miss}$, will provide important complementary tests of possible NP explanations. We therefore look forward to the upcoming Belle~II data, which should definitively confirm or refute the observed excess and, in the process, potentially unveil new degrees of freedom at the GeV scale.

%%%%%%%%%%%%%%%%%%%%%%%%%%%%%%%%%%%%%%%%%%
%
\begin{acknowledgments}
PDB, SF and JFK acknowledge financial support from the Slovenian Research Agency (research core funding No. P1-0035, J1-3013 and N1-0321). MN acknowledges the financial support by the Spanish Government (Agencia Estatal de Investigaci\'on MCIN/AEI/10.13039/501100011033)  and the European Union NextGenerationEU/PRTR through the “Juan de la Cierva” program (Grant No. JDC2022-048787-I) and through Grants No. PID2020-114473GB-I00 and No. PID2023-146220NB-I00. 
\end{acknowledgments}
%
%%%%%%%%%%%%%%%%%%%%%%%%%%%%%%%%%%%%%%%%%%

\newpage

\appendix

\begin{widetext}

%%%%%%%%%%%%%%%%%%%%%%%%%%%%%%%%%%%%%%%%%%
%
\section{Decay Rates}
\label{app:decays}
%
%%%%%%%%%%%%%%%%%%%%%%%%%%%%%%%%%%%%%%%%%%

Firstly, we review the decay rates for the two-body processes $B \to K^{(*)}\sum X$, with $\sum X\in\{\phi, V\}$. While we work in the chiral basis in the main text, we give the following expressions in terms of the parity basis coefficients, with appropriate conversions provided.

The two-body decay rates to a real scalar boson $\phi$ are
\begin{align}
\Gamma(B\to K \phi) &= \frac{|\vec{p}_{K}|}{8\pi} |g_S|^2 \frac{m_B^2\delta_K^2}{(m_b - m_s)^2} f_{0}^2\,, \label{eq:BKphi} \\
\Gamma_T(B\to K^* \phi) &= 0 \,, \label{eq:BKsphiT}\\
\Gamma_L(B\to K^* \phi) &= \frac{|\vec{p}_{K^*}|}{2\pi}|g_P|^2\frac{|\vec{p}_{K^{*}}|^2}{(m_b + m_s)^2} A_{0}^2\,, \label{eq:BKsphiL}
\end{align}
with $\delta_{K^{(*)}} \equiv (1 - m_{K^{(*)}}^2/m_B^2)$, and the form factors $f_0$ and $A_0$ evaluated at $q^2 = m_\phi^2$. The couplings $g_S$ and $g_P$ are given in terms of the chiral basis couplings as
\begin{align}
\begin{pmatrix}
g_{S} \\
g_{P} \\
\end{pmatrix} = 
P\frac{v}{\sqrt{2}\Lambda}
\begin{pmatrix}
C_{d\phi}^{S,L} \\
C_{d\phi}^{S,R} \\
\end{pmatrix} = -P'\frac{i}{\Lambda}\begin{pmatrix}
C_{d\phi}^{V,L} \\
C_{d\phi}^{V,R} \\
\end{pmatrix}\,;\quad P \equiv\frac{1}{2}
\begin{pmatrix}
1 & 1 \\
-1 & 1
\end{pmatrix} \,,\quad P' \equiv\frac{1}{2}
\begin{pmatrix}
m_b - m_s & m_b - m_s \\
-m_b - m_s & m_b + m_s
\end{pmatrix}\,.
\end{align}

The two-body decay rates to a real vector boson $V$ are
\begin{align}
\Gamma(B\to K V) &= \frac{|\vec{p}_{K}|}{2\pi}\bigg[|h_{V}|^2\frac{|\vec{p}_K|^2}{m_V^2} f_+^2 +  4|h_{T}|^2 \frac{m_V^2}{\Lambda^2}\frac{|\vec{p}_K|^2}{(m_B+m_K)^2}f_T^{2} + 4\,\Re[h_V h_T^*]\frac{|\vec{p}_K|^2}{(m_B+m_K) \Lambda}f_+ f_T \bigg]\,, \label{eq:BKV}\\
\Gamma_T(B\to K^* V) &= \frac{|\vec{p}_{K^*}|}{2\pi}\bigg[2|h_{V}|^2 \frac{|\vec{p}_{K^*}|^2}{(m_B + m_{K^*})^2} V^2 + 8|h_{T}|^2\frac{|p_{K^*}|^2}{\Lambda^2} T_{1}^{2} - 8\,\Re[h_V h_T^*]\frac{|\vec{p}_{K^*}|^2}{(m_B + m_{K^*}) \Lambda}V T_1 \nonumber \\
& \hspace{4.0em} + |h_{A}|^2\frac{(m_B + m_{K^*})^2}{2m_B^2} A_{1}^2 + 2|h_{\tilde{T}}|^2 \frac{m_B^2\delta_{K^*}^2}{\Lambda^2} T_{2}^2 - 2\,\Re[h_A h_{\tilde{T}}^*]\frac{(m_B + m_{K^*})\delta_{K^*}}{\Lambda} A_1 T_2\bigg]  \,, \label{eq:BKsVT} \\
\Gamma_L(B\to K^* V) &= \frac{8|\vec{p}_{K^*}|}{\pi}\bigg[|h_{A}|^2\frac{m_{K^*}^2}{m_V^2}A_{12}^2 +|h_{\tilde{T}}|^2 \frac{m_V^2}{\Lambda^2}\frac{m_{K^*}^2}{(m_B + m_{K^*})^2} T_{23}^2 - 2\,\Re[h_A h_{\tilde{T}}^*]\frac{m_{K^*}^2}{(m_B  + m_{K^*})\Lambda}A_{12} T_{23}\bigg]\,, \label{eq:BKsVL}
\end{align}
with the form factors $f_+$, $f_T$, $V$, $A_1$, $A_{12}$, $T_1$, $T_2$ and $T_{23}$  evaluated at $q^2 = m_V^2$. The couplings $h_V$, $h_A$, $h_T$ and $h_{\tilde{T}}$ are converted to the chiral basis couplings as
\begin{align}
\begin{pmatrix}
h_{V} \\
h_{A} \\
\end{pmatrix} = 
P
\begin{pmatrix}
C_{dV}^{V,L} \\
C_{dV}^{V,R} \\
\end{pmatrix}\,, \quad 
\begin{pmatrix}
h_{T} \\
h_{\tilde{T}} \\
\end{pmatrix} = 
P\frac{v}{\sqrt{2}\Lambda}
\begin{pmatrix}
C_{dV}^{T,L} \\
C_{dV}^{T,R} \\
\end{pmatrix}\,,
\end{align}
respectively. The outgoing three-momentum of $K^{(*)}$ in Eqs.~\eqref{eq:BKphi}--\eqref{eq:BKsVL} is given by
\begin{align}
\label{eq:K-3momentum}
|\vec{p}_{K^{(*)}}| = \frac{\lambda^{1/2}(m_B^2,q^2,m_{K^{(*)}}^2)}{2m_B}\,,
\end{align}
with $q^2 = m_X^2$ and $\lambda(x,y,z) = (x - y - z)^2 - 4 y z$ is the K\"{a}ll\'{e}n function.

Next we review the decay rates of the three-body processes $B \to K^{(*)}\sum X$, for $\sum X\in\{\phi\phi, \phi\bar{\phi}, \psi\psi, \psi\bar{\psi}\}$. For the decays to a complex scalar $\phi$, the differential rates are
\begin{align}
\frac{d\Gamma(B\to K \phi\bar{\phi})}{dq^2} &= \frac{\beta_{\phi}}{96\pi^3}\frac{|\vec{p}_{K}|}{\Lambda^2}\bigg[\frac{3}{4}|g_{SS}|^2\frac{m_B^2\delta_{K}^2}{(m_b - m_s)^2} f_{0}^2 + |g_{VV}|^2\frac{|\vec{p}_K|^2}{\Lambda^2} \beta_{\phi}^2  f_{+}^2\bigg]\,,  \label{eq:BK_phiphi} \\
\frac{d\Gamma_T(B\to K^* \phi\bar{\phi})}{dq^2} &= \frac{\beta_{\phi}}{96\pi^3}\frac{|\vec{p}_{K^*}|q^2}{\Lambda^4} \beta_{\phi}^2\bigg[2|g_{VV}|^2 \frac{|\vec{p}_{K^*}|^2}{(m_B + m_{K^*})^2} V^2 + |g_{AV}|^2 \frac{(m_B + m_{K^*})^2}{2m_B^2} A_{1}^2\bigg] \,, \label{eq:BKs_phiphi_T} \\
\frac{d\Gamma_L(B\to K^* \phi\bar{\phi})}{dq^2} &= \frac{\beta_{\phi}}{96\pi^3}\frac{|\vec{p}_{K^*}|}{\Lambda^2}\bigg[3|g_{PS}|^2\frac{|\vec{p}_{K^*}|^2}{(m_b + m_s)^2}A_{0}^2 + 16 |g_{AV}|^2 \frac{m_{K^*}^2}{\Lambda^2} \beta_{\phi}^2 A_{12}^2\bigg] \,, \label{eq:BKs_phiphi_L}
\end{align}
with $\beta_X = \sqrt{1 - 4m_X^2/q^2}$ and the three-momentum of $K^{(*)}$ is given by Eq.~\eqref{eq:K-3momentum}. For a real scalar field, i.e. $\phi = \bar{\phi}$, the contributions from $g_{VV}$ and $g_{AV}$ vanish, and the remaining contributions from $g_{SS}$ and $g_{SP}$ are a factor of $2$ larger at the amplitude level; hence, $g_{SS} \to 2 g_{SS}$ and $g_{PS} \to 2 g_{PS}$ must be made in Eq.~\eqref{eq:BK_phiphi}. The rates should also be multiplied by factor of $1/2$ to account for identical outgoing states. The parity-basis coefficients can be converted to those in the chiral basis as
\begin{align}
\begin{pmatrix}
g_{SS} \\
g_{PS} \\
\end{pmatrix} = 
P\frac{v}{\sqrt{2}\Lambda}
\begin{pmatrix}
C_{d\phi\phi}^{S,L} \\
C_{d\phi\phi}^{S,R} \\
\end{pmatrix}\,, \quad \begin{pmatrix}
g_{VV} \\
g_{AV} \\
\end{pmatrix} = 
P \begin{pmatrix}
C_{d\phi\phi}^{V,L} \\
C_{d\phi\phi}^{V,R} \\
\end{pmatrix}\,.
\end{align}
For the decays to a pair of Dirac fermions $\psi\bar{\psi}$, the relevant differential rates are
\begin{align}
\frac{d\Gamma(B\to K \psi\bar{\psi})}{dq^2} & =  \frac{\beta_{\psi}}{24\pi^3}\frac{|\vec{p}_{K}|q^2}{\Lambda^4}\bigg[\Big(|f_{VV}|^2\beta_\psi^{\prime 2} + |f_{VA}|^2\beta_{\psi}^2\Big)\frac{|\vec{p}_K|^2}{q^2}f_{+}^2 + \frac{3}{2}|f_{VA}|^2\frac{m_\psi^2 m_B^2 \delta_K^2}{q^4} f_{0}^2\bigg] \,, \label{eq:BK_psipsi}\\
\frac{d\Gamma_T(B\to K^{*} \psi\bar{\psi})}{dq^2} &= \frac{\beta_{\psi}}{24\pi^3}\frac{|\vec{p}_{K^*}|q^2}{\Lambda^4}\bigg[\Big(|f_{VV}|^2\beta_\psi^{\prime 2} + |f_{VA}|^2\beta_\psi^2\Big) \frac{2|\vec{p}_{K^*}|^2}{(m_B + m_{K^*})^2} V^{2} \nonumber \\
&\hspace{7.5em}+ \Big(|f_{AV}|^2\beta_\psi^{\prime 2} + |f_{AA}|^2\beta_\psi^{2}\Big) \frac{(m_B + m_{K^*})^2}{2m_B^2} A_{1}^2\bigg] \,, \label{eq:BKs_psipsi_T}\\
\frac{d\Gamma_L(B\to K^{*} \psi\bar{\psi})}{dq^2} &= \frac{\beta_{\psi}}{24\pi^3}\frac{|\vec{p}_{K^*}|q^2}{\Lambda^4}\bigg[ \Big(|f_{AV}|^2\beta_\psi^{\prime 2} + |f_{AA}|^2\beta_\psi^{2}\Big) \frac{16m_{K^*}^2}{q^2}A_{12}^2 + 6|f_{AA}|^2\frac{m_\psi^2 |\vec{p}_{K^*}|^2}{q^4} A_0^{2}\bigg] \,, \label{eq:BKs_psipsi_L}
\end{align}
where $\beta_\psi^{\prime 2} \equiv (3-\beta_\psi^2)/2 = (1 + 2m_\psi^2/q^2)$. We do not consider scalar and tensor couplings in this work; the full expressions including these are given in App.~B of Ref.~\cite{Bolton:2024egx}. The parity basis coefficients in Eqs.~\eqref{eq:BK_psipsi}--\eqref{eq:BKs_psipsi_L} are given by those in the chiral basis as
\begin{align}
\begin{pmatrix}
f_{VV} \\
f_{VA} \\
f_{AV} \\
f_{AA} \\
\end{pmatrix} = 
\frac{1}{4}\begin{pmatrix}
1 & 1 & 1 & 1 \\
-1 & 1 & -1 & 1 \\
-1 & -1 & 1 & 1 \\
1 & -1 & -1 & 1 \\
\end{pmatrix}\begin{pmatrix}
C_{d\psi}^{V,LL} \\
C_{d\psi}^{V,LR}  \\
C_{d\psi}^{V,RL}  \\
C_{d\psi}^{V,RR}  \\
\end{pmatrix} \,. \label{eq:chiraltoparity}
\end{align}
For a pair of invisible Majorana fermions ($\psi = \psi^c$) in the final state, the vector current vanishes identically, $\bar{\psi}\gamma_\mu \psi = 0$, so the replacements $f_{VV,AV} \to 0$ should be made in Eqs.~\eqref{eq:BK_psipsi}--\eqref{eq:BKs_psipsi_L}. An additional contraction is possible at the level of the amplitude for the operators involving the axial vector current $\bar{\psi}\gamma_\mu \gamma_5\gamma$, resulting in $f_{VA,AA} \to 2 f_{VA, AA}$ for the remaining coefficients. For the coefficients in the chiral basis, this is equivalent to $C_{d\psi}^{V,XL} = -C_{d\psi}^{V,XR}$ for $X \in \{L,R\}$. Additionally, a factor of $1/2$ is required to account for identical outgoing states.

Finally, we give expressions for the decay rates of $B_s \to \sum X$ with $\sum X\in\{\phi\phi, \phi\bar{\phi}, \psi\psi, \psi\bar{\psi}\}$. For a pair of outgoing scalars,
\begin{align}
\Gamma(B_s \to \phi\bar{\phi}) = \frac{\beta_\phi}{16\pi}\frac{f_{B_s}^2 m_{B_s}}{\Lambda^2}\bigg[|g_{PS}|^2\frac{m_{B_s}^2}{(m_b + m_s)^2} + \frac{1}{4}|g_{AV}|^2 \frac{m_{B_s}^2}{\Lambda^2
}\bigg]\,,
\end{align}
where $\beta_X = \sqrt{1 - 4m_X^2/m_{B_s}^2}$. If $\phi$ is real, the same modifications as under Eq.~\eqref{eq:BKs_phiphi_L} should be made.

For a pair of outgoing Dirac fermions,
\begin{align}
\Gamma(B_s \to \psi\bar{\psi}) = \frac{\beta_\psi}{2\pi}\frac{f_{B_s}^2 m_{B_s}^3}{\Lambda^4}|f_{AA}|^2\frac{m_\psi^2}{m_{B_s}^2}\,.
\end{align}
Likewise, for Majorana fermions, the replacements under Eq.~\eqref{eq:BKs_psipsi_L} should be made. For the same values of $f_{AA}$ and $m_\psi$, the decay rate to Majorana fermions is a factor of 2 larger than to Dirac fermions.

\end{widetext}

\bibliography{main}

\end{document}